\documentclass[12pt]{iopart}
\usepackage{bm}
\usepackage{hyperref}
\usepackage{amssymb}
\newcommand{\dn}[2]{{\mathrm{d}^{{#1}}{{#2}}}}
\newcommand{\fmeasure}[1]{{[\mathrm{d}{{#1}}]}}
\newcommand{\deriv}[2]{{\frac{\mathrm{d}{{#1}}}{\mathrm{d}{{#2}}}}}

\renewcommand{\d}{\mathrm{d}}

\newcommand{\R}{\mathcal{R}}
\newcommand{\vect}[1]{{\bm{\mathrm{{#1}}}}}
\newcommand{\planckmass}{M_{\mathrm{P}}}
\renewcommand{\e}[1]{{\mathrm{e}^{{#1}}}}
\newcommand{\imag}{{\mathrm{i}}}

\newcommand{\fnl}{f_{\mathrm{NL}}}
\renewcommand{\epsilon}{\varepsilon}
\renewcommand{\phi}{\varphi}
\newcommand{\G}{\mathcal{G}}
\newcommand{\M}{\mathcal{M}}
\newcommand{\N}{\mathcal{N}}
\newcommand{\A}{\mathcal{A}}
\newcommand{\Q}{\mathcal{Q}}
\newcommand{\prob}{\mathbb{P}}

\begin{document}
\title{Primordial non-gaussianities from multiple-field inflation}
\date{\today}
\author{David Seery and James E. Lidsey}
\address{Astronomy Unit, School of Mathematical Sciences\\
  Queen Mary, University of London\\
  Mile End Road, London E1 4NS\\
  United Kingdom}
\eads{\mailto{D.Seery@qmul.ac.uk}, \mailto{J.E.Lidsey@qmul.ac.uk}}
\begin{abstract}
We calculate the three-point correlation function 
evaluated at horizon crossing for a set of 
interacting scalar fields
coupled to gravity during inflation. This provides the initial condition for
the three-point function of the curvature perturbation in
the Sasaki--Stewart $\delta N$ formulation.
We find that the effect is small, of the order of a slow-roll parameter,
and that the non-gaussianity can be determined on large scales once the
unperturbed background evolution is known.
As an example of the use of our formalism, we calculate the
primordial non-gaussianity arising in a model of assisted inflation.
\vspace{3mm}
\begin{flushleft} \textbf{Keywords}: Cosmological perturbation theory,
Inflation,
Cosmology of theories beyond the SM,
Physics of the early universe
\end{flushleft}
\end{abstract}
\maketitle
\section{Introduction}

The inflationary paradigm is an attractive proposal 
for the evolution of the very early universe 
\cite{starobinsky,guth,albrecht-steinhardt,hawking-moss,linde}.
During inflation
the universe underwent a phase of accelerated expansion driven by one or more
self-interacting scalar fields. Despite the 
appealing simplicity behind the central idea of inflation, 
it has proved difficult 
to discriminate between the large number of different models that 
have been developed to date \cite{lyth-riotto}. 
The simplest classes of models typically predict that 
the spectrum of primordial density perturbations 
generated quantum-mechanically during inflation should be 
nearly scale-invariant and Gaussian-distributed. 
(For a review, see, e.g., Ref. \cite{liddle-lyth}.) These generic 
predictions are in good agreement with cosmological 
observations, most notably those arising from measurements of the temperature 
anisotropy and polarization of the Cosmic Microwave Background (CMB)
\cite{bennett-banday,wmap}. Such 
observations lead to strong constraints on theoretical 
model building. 

However, if further progress is to be made in our understanding 
of the microphysics of the early universe, 
it will become increasingly necessary to extend 
the theoretical framework beyond the leading-order effects of scale-invariant, 
Gaussian fluctuations. In particular, deviations away from Gaussian 
statistics represent a potentially powerful discriminant 
between competing inflationary models and have attracted considerable
recent interest
\cite{bartolo-matarrese-review,maldacena-nongaussian,seery-lidsey,
creminelli,lyth-rodriguez,lyth-rodriguez-a,acquaviva-bartolo,
calcagni-nongaussian,rigopoulos-shellard,rigopoulos-shellard-vantent}.
Moreover, measurements of CMB anisotropies are 
now approaching a level of precision where such differences
could soon be detectable and, indeed, the first 
non-gaussian signals in the data may already have
been observed \cite{mcewen-hobson,mukherjee-wang,vielva-martinez-gonzalez,
larsen-wandelt} (see also \cite{aghanim-kunz}).

In general, any Gaussian random variable $X$ with zero mean and variance
$\sigma^2$ has a probability measure given by
\begin{equation}
  \prob \left[ X \in (x, x + \d x) \right] \, \d x
  \propto \exp \left( - \frac{x^2}{2\sigma^2} \right) \, \d x ,
\end{equation}
and the correlation functions of $X$ have the form
\begin{equation}
  \langle X^n \rangle = \int x^n \;
  \prob\left[ X \in (x, x + \d x) \right] \, \d x .
\end{equation}
The rules of Gaussian integration imply that $\langle X^n \rangle$ vanishes
whenever $n$ is odd, whereas $\langle X^n \rangle$ can be written in terms of
$\langle X^2 \rangle$ for even $n$.  Accordingly, we expect violations
of Gaussian statistics to manifest themselves as deviations from these
simple rules.  The same principle holds when dealing with
fluctuations $\delta \phi$ in some quantum field, where the 
semiclassical probability density is given by $\prob(\delta \phi) 
\sim \e{-I[\delta \phi]}$ and
$I$ is the Euclidean action. When $I$ describes a free theory with no
interactions, $\delta \phi$ may be viewed as a
Gaussian random variable. However, if 
interactions are present, deviations from pure Gaussian statistics will 
arise. More specifically, if $\delta\phi$ 
represents an `almost Gaussian' random variable, then it is 
expected (in a largely model-independent sense) 
that the three-point function $\langle
\delta \phi^3 \rangle$ will provide the largest contribution 
to the non-gaussianity. 

In typical single- or multi-field inflationary
models, the level of non-gaussianity can be calculated analytically
\cite{bartolo-matarrese-review}. 
The single-field scenario has been well-studied
\cite{falk-rangarajan,gangui-lucchin,maldacena-nongaussian,acquaviva-bartolo,
bartolo-matarrese-review}, 
and some effort has focused recently on non-standard 
theories of gravity, in the hope of identifying observational signatures 
that differ from those of canonical inflation
based on the Einstein action
\cite{alishahiha-silverstein,arkani-hamed-creminelli,creminelli,
calcagni-nongaussian}.
Some earlier investigations considered non-gaussian 
fluctuations, but were specifically directed at
the formation of primordial black holes during the final stages 
of inflation, when the slow-roll approximation
has broken down \cite{bullock-primack,ivanov}.
In general, however,
the superhorizon evolution of the curvature perturbation 
during single-field inflation is
simple, and there is a robust prediction that the level of primordial
non-gaussianity as measured by the 
three-point correlation function 
cannot be large \cite{maldacena-nongaussian}, even allowing
for exotic effects such as higher-derivative operators of the 
scalar field or a speed of sound that is  
different from that of light \cite{seery-lidsey,creminelli}.

After the
end of inflation, one must evolve this small non-gaussianity through the
epochs of radiation and matter domination in order to arrive at the
temperature anisotropy $\delta T/T$ which is seen on the microwave background
sky \cite{bartolo-matarrese-inflation,bartolo-matarrese-fnl}.  This evolution
after the end of inflation gives an additional source of non-gaussianity
which is comparable to the primordial non-gaussianity in the case of
single-field inflation.

By comparison, our understanding 
of the multi-field scenario is less developed, although 
some progress has been made
\cite{bento-bertolami-a,bento-bertolami-b,sakharov-khlopov,
sakharov-sokoloff,enqvist-jokinen-a,enqvist-jokinen-b,
rigopoulos-shellard,rigopoulos-shellard-vantent,lyth-rodriguez,
lyth-rodriguez-a,enqvist-vaihkonen,enqvist-jokinen,bernardeau-uzan}.
Typically, the evolution of the curvature
perturbation on superhorizon scales implies that a significant non-gaussian
signal can in principle be generated during inflation.
Indeed,
a fairly general prescription for calculating the three-point function of
fluctuations
was recently presented \cite{lyth-rodriguez-a}, in which
it was assumed that the
curvature perturbation evolves substantially outside the horizon due
to pressure effects after the end of inflation.
Such superhorizon evolution effects generically arise 
when more than one fluid is cosmologically important,
or when a single fluid is present
but with an indefinite equation of state. 
In the multiple-field case, therefore, the primordial
three-point function consists of a microphysical component, 
calculated from the
properties of the matter fluctuations around the epoch of horizon exit, 
and a superhorizon component, which is determined 
purely by large-scale gravitational effects.
If the superhorizon component becomes large, which typically will not
happen until after the end of inflation,
it may dominate the primordial non-gaussianity, in which case it is 
consistent to neglect the microphysical contribution. This actually
occurs, for example, in the curvaton scenario \cite{lyth-wands,
moroi-takahashi,enqvist-sloth,lyth-rodriguez,lyth-rodriguez-a}.
In this respect, the theory has some
similarities with the stochastic approach to inflation, which has been revived
recently within the context of 
non-gaussianity \cite{rigopoulos-shellard,rigopoulos-shellard-vantent}. 

However, in the absence of any estimate for the 
microphysical contribution to the three-point 
function, the above prescription is necessarily incomplete, since 
it is not possible to verify explicitly that the superhorizon
contribution dominates.  
In this paper, we supply the missing ingredient by
calculating the three-point function evaluated at the epoch 
of horizon crossing for a generic set of $\N$ scalar fields $\{ \phi^I \}$ 
coupled to gravity. We consider a (Lorentzian) action of the form 
\begin{equation}
  \label{intro:action}
  S = \frac{1}{2} \int \dn{4}{x} \; \sqrt{-g} \, \left[ R -
  \G_{IJ} \, \partial_{\mu} \phi^I \partial^{\mu} \phi^J - 2V(\phi) \right] .
\end{equation}
where the interaction potential $V(\phi)$ 
is assumed to be an arbitrary function of the fields, 
$\phi^I$, and $\G_{IJ}$ represents the metric
on the manifold parametrized by the scalar field values. 
(We refer to this metric as the
`target space' metric).  The action for $\N$
canonically normalized scalar fields minimally coupled to gravity 
is recovered by specifying $\G_{IJ} = \delta_{IJ}$, 
where $\delta_{IJ}$ represents the Kronecker-delta. More generally,
the kinetic sector of \eref{intro:action} 
may be invariant under a global symmetry 
group $J$, in which case the scalar fields parametrize the coset space 
$J/K$ with a line element given by $\d s^2 = G_{IJ}\, \d\phi^I \d\phi^J$, 
where $K$ is the maximal compact subgroup of $J$. 
In such models, the metric components generically 
exhibit a direct dependence on the scalar fields $\phi^I$.
In this paper, we restrict our attention
to the case where the target space is approximately flat,
at least over the region dynamically explored by the fields $\phi^I$,
and the metric can be
brought to the field-\emph{independent} form $\G_{IJ}=\delta_{IJ}$ by
an appropriate choice of parametrization.  On the other hand, we
allow the potential $V(\phi)$ to be quite arbitrary,
subject to the usual slow-roll conditions. 
This choice of action
covers a wide class of multi-field inflationary scenarios. 

The outline of the paper is as follows. In Section~\ref{sec:mfield}, we
briefly describe how to calculate the primordial 
non-gaussianity produced during multi-field inflation, as measured by
the three-point function of the curvature perturbation. 
The crucial ingredients for the calculation are: (i)
an expression for the curvature perturbation itself,
which is provided by the Sasaki--Stewart $\delta N$ formalism 
\cite{sasaki-stewart},
and: (ii) an estimate of the scalar three-point
function just after horizon exit. This three-point function is 
derived in Section~\ref{sec:threept} and is 
presented in Eqs. (\ref{threept:result})--(\ref{threept:a}). 
It represents the central result of the paper. In Section~\ref{sec:single},  
we verify that our formalism reproduces the
well-known result of Maldacena \cite{maldacena-nongaussian}
when specialised to the case of a single scalar field. 
In Section~\ref{sec:assisted} the non-gaussianity generated 
during assisted inflation
\cite{liddle-mazumdar,barreiro-copeland,green-lidsey} is calculated
as an example of our method.
Finally, we conclude in Section~\ref{sec:conclusions}.
 
Units are chosen such that $c = \hbar = \planckmass = 1$, where
$\planckmass = (8 \pi G)^{-1/2}$ represents the reduced Planck mass.

\section{Non-gaussianity from multiple fields}
\label{sec:mfield}

\subsection{The background model}

We assume throughout that the background model corresponds to the 
spatially flat Friedmann--Robertson--Walker (FRW) universe, with a 
metric 
\begin{equation}
  \d s^2 = -\d t^2 + a^2 \delta_{ij} \d x^i \d x^j .
\end{equation}
The scalar field equations take the form of a set of coupled 
Klein--Gordon equations
\cite{sasaki-stewart,hwang-noh-multi,nibbelink-vantent,van-tent}: 
\begin{equation}
\label{scalareom}
  \ddot{\phi}^I + 3 H \dot{\phi}^I + \G^{IJ} V_{,J} = 0 ,
\end{equation}
where $H = \dot{a}/a$ is the Hubble parameter, $a$ is the scale factor, 
$\dot{\phi}^I = \d \phi^I / \d t$ and a comma denotes 
partial differentiation. 

The gravitational equation of motion is
\begin{equation}
\label{Hdot}
  2 \dot{H} + 3 H^2 = - \frac{1}{2} \G_{IJ} \dot{\phi}^I \dot{\phi}^J
  + V(\phi) ,
\end{equation}
together with the Friedmann constraint equation
\begin{equation}
\label{F}
  3H^2 = \frac{1}{2} \G_{IJ} \dot{\phi}^I \dot{\phi}^J + V(\phi) .
\end{equation}
Eqs. (\ref{Hdot}) and (\ref{F}) together imply that 
\begin{equation}
\label{Ray}
  \dot{H} = - \frac{1}{2} \G_{IJ} \dot{\phi}^I \dot{\phi}^J  .
\end{equation}
There is also an analogue of the single-field Hamilton--Jacobi equation,
which reads
\begin{equation}
  \label{intro:hj}
  H_{,I} = - \frac{1}{2} \dot{\phi}_I .
\end{equation}

As in single-field inflation, it will prove convenient to 
introduce a set of slow-roll parameters. One such parameter 
is the slow-roll matrix $\epsilon^{IJ}$: 
\begin{equation}
\label{epsilonIJ}
  \epsilon^{IJ} = \frac{\dot{\phi}^I \dot{\phi}^J}{2H^2} =
  2 \G^{IM} \G^{JN} \frac{H_{,M} H_{,N}}{H^2} = \epsilon^I \epsilon^J ,
\end{equation}
where we have used \eref{intro:hj}, and
$\epsilon^I$ is the vector
\begin{equation}
  \epsilon^I = \frac{\dot{\phi}^I}{\sqrt{2}H} =
  \frac{\sqrt{2} \G^{IJ} H_{,J}}{H} .
\end{equation}
The standard slow-roll parameter, $\epsilon = - \dot{H}/H^2$, can then be
expressed in terms of $\epsilon^{IJ}$, since
$\epsilon = \tr \epsilon^{IJ} = \G_{IJ} \epsilon^{IJ}$. As a result, we
generally expect $|\epsilon^{IJ}| \ll 1$ whenever slow-roll is valid
except in models where finely-tuned cancellations between the 
components of $\epsilon^{IJ}$ may occur.  Specifically, for generic
theories which are not tuned to have anomalous parameter values, one should
expect
\begin{equation}
  \epsilon^{IJ} = \Or\left( \frac{\epsilon}{\N} \right),
  \quad
  \epsilon^I = \Or\left( \frac{\epsilon^{1/2}}{\N^{1/2}} \right) .
\end{equation}

The time derivative of $\epsilon^{IJ}$ is determined by 
\begin{equation}
  \dot{\epsilon}^{IJ} = 2 \epsilon H \left( \epsilon^{IJ} - \eta^{IJ} \right),
\end{equation}
where we have defined a second slow-roll matrix 
\begin{equation}
\label{etaIJ}
  \eta^{IJ} = \frac{\ddot{\phi}^I \dot{\phi}^J + \dot{\phi}^J \ddot{\phi}^J}
  {4 H \dot{H}} .
\end{equation}
This matrix generalizes the slow-roll parameter $\eta$ of single-field
inflation, as can be deduced by substituting in Eq. (\ref{Ray}) to 
yield  $\eta^{\phi\phi} = - \ddot{\phi}/H\dot{\phi} = \eta$.  

Finally, we introduce a third matrix defined by
\begin{equation}
\label{tildeetaIJ}
  \tilde{\eta}_{IJ} = \frac{V_{,IJ}}{3H^2} .
\end{equation}
This is related to $\epsilon^{IJ}$ and $\eta^{IJ}$ by the simple rule
\begin{equation}
  \label{threept:etatilde}
  \epsilon^{IJ} + \eta^{IJ} = \tilde{\eta}_{MN}
  \frac{\G^{M(I} \epsilon^{J)N}}{\epsilon} .
\end{equation}
The parameter $\tilde{\eta}_{IJ}$ represents a generalization of the slow-roll
parameter $V''/V$ of single-field inflation, and can be related to
the alternative definition $-\ddot{\phi}/H\dot{\phi}$.  In multi-field 
models, however, the index arrangement in \eref{threept:etatilde} means
that it is convenient to maintain two separate definitions. 

We note that Eq. \eref{threept:etatilde} implies 
that $\eta^{IJ}$ and $\tilde{\eta}_{IJ}$
are of the same order in slow-roll. The utility of this formulation
of the slow-roll approximation is essentially restricted
to the case of a flat target space.

It will also be useful to have on hand an explicit expression for the
number of e-foldings in this model.  
One defines the number of e-folds $N$ which
elapse between some initial value $a_i$ of the scale factor and a final value
$a_f$ as $N = \ln a_f/a_i$, so
\begin{equation}
  \label{intro:ndef}
  \d N = H \, \d t = - \frac{1}{\epsilon} \; \d \ln H =
  - \frac{\epsilon_I}{\sqrt{2} \epsilon} \; \d \phi^I .
\end{equation}

\subsection{The uniform density curvature perturbation}
\label{sec:zeta}

We now consider perturbation theory around 
the above homogeneous background. Since it is assumed that 
the scalar fields $\phi^I$ dominate the energy density of
the universe, any perturbations $\delta \phi^I$ in these fields
necessarily produce a disturbance in the energy--momentum tensor, which
back-reacts on the spacetime curvature.
Therefore, when working with the perturbation theory of these scalar fields,
we must also take into account the effect of scalar metric
fluctuations in order to include the effect of this back-reaction.

To first-order in perturbation theory, these metric fluctuations
can be written as \cite{mfb}
\begin{equation}
  \label{mfield:metric}
  \d s^2 = -(1+2\Phi) \d t^2 + 2 a^2 B_{,i} \, \d x^i
  \d t + a^2 \left[ (1 - 2 \Psi) \delta_{ij} + 2 E_{,ij} \right]
  \d x^i \d x^j .
\end{equation}
Not all of these
degrees of freedom are dynamically important, however, since 
they are related by the gauge transformations
$t \mapsto t + \delta t$ and $x^i \mapsto x^i + \delta x^i$, which reshuffle
$\Phi$, $\Psi$, $B_{,i}$ and $E_{,ij}$.  

To obtain a gauge-invariant
measure of the strength of such fluctuations, it is sufficient 
to consider scalar curvature invariants of the 
three-dimensional hypersurfaces $t= \mbox{constant}$. 
The simplest such invariant is given by 
\begin{equation}
  \Psi = \frac{a^2}{4} \partial^{-2} R^{(3)} ,
\end{equation}
where $R^{(3)}$ is the three-dimensional Ricci scalar on the spatial
hypersurfaces, and $\partial^{-2}$ is the inverse 
Laplacian\footnote{This is most simply defined in Fourier space, 
where the Laplacian
$\partial^2$ is represented as multiplication by $-k^2$.  Similarly, the
inverse Laplacian corresponds to multiplication by $-k^{-2}$.}.
In single-field models, this quantity has an attractive interpretation.
Under a temporal
gauge transformation $t \mapsto t + H \delta t$, the field $\Psi$
transforms according to the rule
\begin{equation}
  \label{mfield:psixfm}
  \Psi \mapsto \Psi + H \delta t .
\end{equation}
Hence, the curvature scalar on comoving hypersurfaces
(defined by the condition that the scalar field perturbation 
$\delta\phi$ vanishes) is given by 
\begin{equation}
  \R = \Psi_{\delta \phi = 0} = \Psi + H \frac{\delta\phi}{\dot{\phi}} .
\end{equation}

One may also consider the curvature scalar on 
uniform density hypersurfaces, defined by $\delta\rho = 0$. 
This scalar is denoted by $\zeta$, where 
$|\zeta| = |\Psi_{\delta \rho = 0}|$ \cite{wands-malik}, but there is no
consistent agreement on sign conventions.
For adiabatic perturbations generated during single-field inflation, the two
sets of hypersurfaces coincide \cite{wands-malik}, 
so modulo choices for signs, it follows that $\R = |\zeta|$.
One can freely work in either the comoving or uniform
density gauges, and the gauge-invariant variable $\R = |\zeta| = 
\Psi + H \delta \phi/\dot{\phi}$
is the only degree of freedom in the theory. This variable
mixes the fluctuations arising from the metric and the scalar field, neither
of which has a separate gauge-invariant interpretation. 
Consequently, at the quantum level one considers the creation of 
$\R$- or $\zeta$-particles from the vacuum rather than that of 
inflaton particles or gravitons.

After horizon exit, $\R$ asymptotes towards a time-independent quantity 
and may be viewed as a function only of position. 
It can therefore be absorbed by an appropriate choice of coordinates 
on the spatial hypersurfaces.  This implies 
that distant regions of the universe differ from each other 
only by how much one has expanded relative to
the other.  Since this is a symmetry of the equations of motion,
$\R$ is conserved on superhorizon scales \cite{maldacena-nongaussian}.

In multi-field models, the situation is not so simple.  It is still
convenient to work with the uniform density
curvature perturbation, $\zeta$, which on
large scales is still equivalent to the comoving curvature perturbation $\R$.
However, expressing $\zeta$ as a mixture of the
metric perturbation $\Psi$ and the scalar field perturbations $\delta \phi^I$
is much more involved. 
This was first achieved at a linear level
by Sasaki \& Stewart \cite{sasaki-stewart}, who
used \eref{mfield:psixfm} to express the difference between an initial
uniform curvature hypersurface and a final comoving hypersurface as
\begin{equation}
  \label{ss}
  \zeta = | H \, \d t | = \d N ,
\end{equation}
where $N$ is the integrated number of e-folds, as defined in \eref{intro:ndef}.
On sufficiently large scales
we expect (given suitable assumptions about the dynamical behaviour
which permit us to ignore time derivatives of the perturbations)
that each horizon volume will evolve as if it were a self-contained universe,
and it therefore follows that (as shown beyond linear order by
Lyth, Malik \& Sasaki \cite{lyth-malik})
\begin{equation}
  \label{mfield:zeta}
  \zeta = N_{,I} \, \delta\phi^I + \frac{1}{2} N_{,IJ}
  \, \delta\phi^I \delta\phi^J + \cdots ,
\end{equation}
where the $\delta\phi^I$ express the deviations of the fields from their
unperturbed values in some given region of the universe, and we employ
the standard summation convention for the scalar field 
indices $I$, \ldots, $J$.
It is this expression for $\zeta$ which was used by Lyth \& Rodriguez
to express the primordial non-gaussianity of $\zeta$ in multi-field
models \cite{lyth-rodriguez-a}.  (See also \cite{boubekeur-lyth}.)

When more than one field is dynamically important,
Eq.~\eref{mfield:zeta} allows $\R$ to evolve outside the horizon
while inflation is still occurring\footnote{To be clear, 
we should point out that there is evolution of the
metric perturbations after the end of inflation.
One source of evolution, arising from the evolution of perturbations
during the epochs of
radiation and matter domination which follow reheating
\cite{bartolo-matarrese-inflation,bartolo-matarrese-fnl}, is
universal to all models of inflation, irrespective of whether many fields
are important or only one. Additionally, in multiple field models,
$N$ will typically
continue to vary after the end of inflation. Since \eref{mfield:zeta} is
still valid, this means that $\zeta$ will continue to change. In this
paper, we only evaluate $\zeta$ up to the end of inflation.},
in contrast with the single-field case.  It can be shown that the
evolution of $\R$ is sourced by a non-adiabatic pressure perturbation
\cite{wands-malik}:
\begin{equation}
  \dot{\R} \simeq -\frac{H}{p + \rho} \delta p_{\mbox{\scriptsize nad}} ,
\end{equation}
where $\delta p_{\mbox{\scriptsize nad}}$ can be written in the form 
\begin{equation}
  \delta p_{\mbox{\scriptsize nad}} = \dot{p} \Gamma = \dot{p}
  \left( \frac{\delta p}{\dot{p}} - \frac{\delta \rho}{\dot{\rho}} \right) ,
\end{equation}
and $\Gamma$ represents the so-called
isocurvature perturbation that measures the displacement
between hypersurfaces of constant density and of constant pressure.  (Note
that $\Gamma$ vanishes whenever $p$ is a definite function of $\rho$.)
Thus, whenever isocurvature perturbations are present, which are
essentially inevitable in a general multi-field scenario, $\R$ will evolve
outside the horizon while inflation is still occuring,
and this evolution must be taken into account in the inflationary
prediction \cite{gordon-wands,nibbelink-vantent}.  The Sasaki--Stewart
formalism conveniently handles these complexities for us.

\subsection{The three-point function in multiple-field models}
\label{sec:mfield-model}

In the remainder of this section, we briefly review
the results of \cite{lyth-rodriguez-a}.  

In order to connect with observations, 
any prediction for the non-gaussianity measured by the three-point
function must be expressed 
in terms of an experimentally relevant parameter.  A common
choice is the non-linearity parameter, $\fnl$, which expresses the departure
of $\zeta$ from a Gaussian random variable:
\begin{equation}
  \label{fnldef}
  \zeta = \zeta_g - \frac{3}{5} \fnl \star \zeta_g^2 ,
\end{equation}
where $\zeta_g$ is Gaussian and we have written a convolution product
since $\fnl$ is typically momentum dependent (and defined by
multiplication in Fourier space \cite{komatsu-spergel,verde-wang}).  We
have dropped an uninteresting zero mode in \eref{fnldef}, which only serves
to fix $\langle \zeta \rangle = 0$.  This so-called primordial non-linearity
is processed into an observable non-linearity of temperature anisotropies
$\fnl^{\delta T/T}$ as described in \cite{bartolo-matarrese-fnl}.

To relate $\fnl$ to the three-point function, we follow Maldacena
\cite{maldacena-nongaussian} and write the latter as
\begin{equation}
  \langle \zeta(\vect{k}_1) \zeta(\vect{k}_2) \zeta(\vect{k}_3)
  \rangle = (2\pi)^3 \delta(\vect{k}_1 + \vect{k}_2 + \vect{k}_3)
  \frac{4\pi^4}{\prod_i k_i^3} | \Delta^2_{\zeta} |^2 \A ,
\end{equation}
where $\A$ is a momentum-dependent function that can be calculated
explicitly, and $\Delta^2_{\zeta}$ is the dimensionless
power spectrum of $\zeta$, defined in the standard way by 
\begin{equation}
  \langle \zeta(\vect{k}_1) \zeta(\vect{k}_2) \rangle =
  (2\pi)^3 \delta(\vect{k}_1 + \vect{k}_2) P_{\zeta}(k_1) ,
  \qquad 
  \Delta^2_{\zeta}(k) = \frac{k^3}{2\pi^2} P_{\zeta}(k) ,
\end{equation}
such  that $\langle \zeta\zeta \rangle = \int_0^{\infty}
\Delta^2_{\zeta}(k) \, \d \ln k$. 
The non-linearity parameter can now be written in the succinct 
form \cite{maldacena-nongaussian}\footnote{There are 
other relations between $\fnl$ and the three-point
function which are common in the literature, usually written in terms
of the bispectrum, but they are all equivalent
to this one.  In the present case, the use of the dimensionless power
spectrum is especially convenient because it will allow us to factorize
the index structure in the multiple-field three-point functions.}:
\begin{equation}
  \label{mfield:fnl}
  \fnl = - \frac{5}{6} \frac{\A}{\sum_i k_i^3} .
\end{equation}

It only remains to use \eref{mfield:zeta} to relate $\fnl$ and the
scalar field three-point functions.
We show in Section~\ref{sec:threept} that the two-point function
for the scalars satisfies
\begin{equation}
  \label{mfield:two}
  \langle \delta \phi^I(\vect{k}_1) \delta\phi^J(\vect{k}_2) \rangle =
  (2\pi)^3 \delta(\vect{k}_1 + \vect{k}_2) \frac{2\pi^2}{k_1^3}
  \Delta^2_{\star} \G^{IJ} ,
\end{equation}
and that the three-point function can be written in the form 
\begin{equation}
  \label{mfield:three}
  \langle \delta \phi^I(\vect{k}_1) \delta \phi^J(\vect{k}_2)
  \delta \phi^K(\vect{k}_3) \rangle = (2\pi)^3 \delta(\vect{k}_1 +
  \vect{k}_2 + \vect{k}_3 \rangle \frac{4\pi^4}{\prod_i k_i^3}
  |\Delta^2_{\star}|^2 \A^{IJK} ,
\end{equation}
where $\Delta^2_{\star}$ is the spectrum of a massless scalar field in
de Sitter space and $\A^{IJK}$ is a momentum-dependent 
function given in Eq. (\ref{threept:a}). The determination of 
$\A^{IJK}$ represents the principal result of this paper.  

Eq. \eref{mfield:zeta} implies that the power spectrum for  
$\zeta$ is given by \cite{sasaki-stewart}
\begin{equation}
  \Delta^2_{\zeta} = \Delta^2_{\star} N_{,I} N_{,J} \G^{IJ} .
\end{equation}
On the other hand,
the three-point function which follows from \eref{mfield:zeta} 
can be written as 
\begin{eqnarray}
 \fl\nonumber
  \langle \zeta(\vect{k}_1) \zeta(\vect{k}_2) \zeta(\vect{k}_3) \rangle =
  N_{,I} N_{,J} N_{,K} \langle \delta \phi^I(\vect{k}_1) \delta \phi^J
  (\vect{k}_2) \delta \phi^K(\vect{k}_3) \rangle \\ \label{mfield:threept}
  \mbox{} + \frac{1}{2}
  N_{,I} N_{,K} N_{,MN} \langle \delta \phi^I(\vect{k}_1)
  \delta \phi^K(\vect{k}_2) [\delta\phi^M \star
  \delta\phi^N](\vect{k}_3) \rangle + \mbox{perms} + \cdots ,
\end{eqnarray}
where $\star$ denotes the convolution product, and for brevity
we have omitted the remaining cross terms in the expansion of $\zeta^3$.
(In particular, there will be terms involving the
product of a spectrum and a bispectrum, of the form
\begin{equation}
  \frac{1}{4} N_{,I} N_{,KL} N_{,MN} \langle
  \delta \phi^I(\vect{k}_1)
  [ \delta \phi^K \star \delta \phi^L ] (\vect{k}_2)
  [ \delta \phi^M \star \delta \phi^N ] (\vect{k}_3) \rangle
  + \mbox{perms} ,
\end{equation}
and a term involving the product of the spectra, of the form
\begin{equation}
  \frac{1}{8} N_{,IJ} N_{,KL} N_{,MN} \langle
  [ \delta\phi^I \star \delta\phi^J ] (\vect{k}_1)
  [ \delta\phi^K \star \delta\phi^L ] (\vect{k}_2)
  [ \delta\phi^M \star \delta\phi^N ] (\vect{k}_3) \rangle .
\end{equation}
The first of these was ignored by Lyth \& Rodriguez
\cite{lyth-rodriguez-a}, because it is proportional
to the three-point function for the $\delta\phi^I$, which these authors
assumed to vanish. These terms can all be treated similarly
\cite{boubekeur-lyth,lyth-rodriguez-a,lyth-axion}
when
included in $\fnl$, but some of the intermediate expressions can become
rather long. For these reason we suppress these higher contributions.)

The first set of terms in Eq. (\ref{mfield:zeta}) 
is just the connected scalar three-point function, since we have 
assumed that all tadpole contributions vanish.
This is mandatory, and equivalent to perturbative stability of the vacuum.
Under the same assumptions, the second group of terms can be reduced to
products of two-point functions. Consequently, by employing 
\eref{mfield:two}--\eref{mfield:three}, Eq. (\ref{mfield:zeta}) 
can be expressed in the form 
\begin{equation}
  \label{mfield:cfn}
  \langle \zeta(\vect{k}_1) \zeta(\vect{k}_2) \zeta(\vect{k}_3) \rangle =
  (2\pi)^3 \delta(\vect{k}_1 + \vect{k}_2 + \vect{k}_3)
  \frac{4\pi^4}{\prod_i k_i^3} | \Delta^2_\star |^2 \A_{\zeta} ,
\end{equation}
where $\A_{\zeta}$ is defined by 
\begin{equation}
  \A_{\zeta} = \A^{IJK} N_{,I} N_{,J} N_{,K} + \G^{IM} \G^{KN}
  N_{,I} N_{,K} N_{,MN} \sum_i k_i^3 .
\end{equation}
Our expression for $\fnl$, Eq. \eref{mfield:fnl}, now implies that the
non-linearity in this model, to a good approximation, is
given by 
\begin{equation}
  \label{fnl}
  \fnl = - \frac{5}{6} \frac{\A^{IJK} N_{,I} N_{,J} N_{,K}}
  {(N_{,I} N_{,J} \G^{IJ})^2 \sum_i k_i^3} -
  \frac{5}{6} \frac{\G^{IM} \G^{KN} N_{,I} N_{,K} N_{,MN}}
  {(N_{,I} N_{,J} \G^{IJ})^2} + \cdots ,
\end{equation}
where by `$\cdots$' we denote the remaining cross terms from
\eref{mfield:zeta} which we did not display.
In principle there will also be other terms coming from the Taylor
expansion of $\zeta$ (in Eq. \eref{mfield:zeta}) beyond second order,
but we assume that the series converges sufficiently fast
that the effect from these higher-order terms is small.
Nevertheless it should be borne in mind that they are
present in principle, and in certain models their contribution could be
significant.
In order to express $\fnl$ as in \eref{fnl}, we have exchanged
$\Delta^2_{\star}$ for $\Delta^2_{\zeta}$ in \eref{mfield:cfn};
this has the effect of introducing the terms involving $N_{,I} N_{,J} \G^{IJ}$
in the denominators of Eq. \eref{fnl}.
We note that
the elements of this expression were first given in \cite{lyth-rodriguez-a}.

Generally speaking, we expect both displayed terms in \eref{fnl} to be of
roughly comparable importance at the end of inflation,
although examples of models where one term is
dominant do exist. For example, assisted inflation represents a 
model where the second
term vanishes, so $\fnl$ arises entirely from the first term.
On the other hand, the authors of Ref.
\cite{lyth-rodriguez-a} focused on the case where the second term is dominant.
Note that the first term and any higher cross-terms containing the
bispectrum $\A^{IJK}$ are
strongly momentum-dependent, whereas the second term and any higher
cross-terms which do not include the bispectrum are
momentum-independent, up to mild scale-dependences which arise from
renormalization effects \cite{boubekeur-lyth,lyth-axion}.

We proceed in the following section to 
calculate the three-point function for the multi-field 
action \eref{intro:action}.

\section{The scalar three-point function}
\label{sec:threept}

\subsection{Perturbations in the uniform curvature gauge}

The first step in determining perturbations in multi-field 
inflation is to select a gauge. The discussion of the previous section,
where \eref{ss} was used to express the curvature perturbation $\zeta$,
implies that the relevant gauge should be 
the uniform curvature slicing, where coordinate time $t$
is chosen in such a way that slices of constant $t$ have zero Ricci
curvature. Such a metric can be written in the ADM form \cite{salopek-bond}:
\begin{equation}
  \label{threept:metric}
  \d s^2 = - N^2 \d t^2 + h_{ij}(\d x^i + N^i \d t)(\d x^j + N^j \d t) .
\end{equation}
This version of the metric incorporates the scalar perturbations $\Phi$
and $B_{,i}$ written down in \eref{mfield:metric} within the 
lapse, $N$,  and  shift, $N^i$. These latter two quantities 
appear as Lagrange multipliers in the action and 
for this reason it is considerably easier to 
perform calculations with \eref{threept:metric} than with the
equivalent expression \eref{mfield:metric}.
The $E_{,ij}$ term of \eref{mfield:metric}
has been removed by a coordinate redefinition. 

One specifies 
$h_{ij} = a^2(t) \delta_{ij}$ in order to obtain spatially flat slices.
The scalar fields on these flat hypersurfaces are then 
written as $\phi^I = \phi^I_0 +
Q^I$, where $\phi^I_0$ are the spatially-homogeneous background values 
of the fields and $Q^I$ represent small perturbations. In these 
coordinates the action (\ref{intro:action}) becomes
\begin{equation}
\label{actionint}
  \fl
  S = - \frac{1}{2} \int N \sqrt{h} \, \left( \G_{IJ} h^{ij} \partial_i \phi^I
  \partial_j \phi^J - 2 V(\phi) \right) +
  \frac{1}{2} \int N^{-1} \sqrt{h} \, \left( E_{ij} E^{ij} - E^2 +
  \G_{IJ} v^I v^J \right) ,
\end{equation}
where $v^I = \dot{\phi}^I - N^j \partial_j \phi^I$.  

There are two constraint equations arising from this action.  
The first follows by demanding that the action be
stationary under variations of $N$:
\begin{equation}
  - \G_{IJ} h^{ij} \partial_i \phi^I \partial_j \phi^J - 2V - \frac{1}{N^2}
  \left( E_{ij} E^{ij} - E^2 + \G_{IJ} v^I v^J \right) = 0 .
\end{equation}
The second constraint arises from the variation in $N^i$ and yields
\begin{equation}
  \left( \frac{1}{N} [ E^j_i - E \delta^j_i ] \right)_{|j} =
  \frac{1}{N} \G_{IJ} v^I \partial_i \phi^J ,
\end{equation}
where $|$ denotes the covariant derivative compatible with 
the spatial metric $h_{ij}$.  These
constraints can be solved to first order by specifying
\begin{eqnarray}
  \label{threept:constr-def}
  N = 1 + \frac{1}{2H} \G_{IJ} \dot{\phi}_0^I Q^J,
\nonumber \\
  N_i = \partial^{-2} \psi = \frac{a^2}{2H} \partial^{-2} \G_{IJ} \left(
  Q^I \ddot{\phi}_0^J - \dot{\phi}_0^I \dot{Q}^J - \frac{\dot{H}}{H}
  \dot{\phi}_0^I Q^J \right) .
\end{eqnarray}
It turns out that the higher-order pieces in $N$ and $N^i$ are not required,
since they cancel out of the second- and third-order terms in the action 
(\ref{actionint}) \cite{maldacena-nongaussian,seery-lidsey}.
In what follows, we will have no need to refer to the total field
$\phi^I = \phi^I_0 + Q^I$, and for notational convenience we therefore 
drop the subscript `0' on $\phi^I_0$ and simply identify 
$\phi^I$ as the homogeneous background field. 

\subsection{The second-order theory}

After integrating by parts in the action \eref{actionint} and employing 
the background field equations \eref{scalareom}--\eref{Ray} 
to simplify the result, we obtain the second-order action
\cite{sasaki-stewart,hwang-noh-multi}: 
\begin{equation}
  \label{S2}  
  S_2 = \frac{1}{2} \int \d t \, \dn{3}{x} \;
  a^3 \left( \G_{IJ} \dot{Q}^I \dot{Q}^J -
  \frac{1}{a^2} \G_{IJ} \partial Q^I \partial Q^J - \M_{IJ} Q^I Q^J
  \right) ,
\end{equation}
where a shorthand notation $\partial Q^I \partial Q^J$ is adopted 
for the scalar product $\delta^{ij} \partial_i Q^I \partial_j Q^J$, and
the mass-matrix, $\M_{IJ}$, is defined by 
\begin{equation}
  \label{defMIJ}
  \M_{IJ} = V_{,IJ} - \frac{1}{a^3} \deriv{}{t} \left( \frac{a^3}{H}
  \dot{\phi}_I \dot{\phi}_J \right) .
\end{equation}
Eq. \eref{S2}  is exact and is valid for 
arbitrary scalar field dynamics.  After changing to conformal time,
$\eta = \int dt/a(t)$, and introducing the canonical
Mukhanov variable $u^I = a Q^I$, one obtains a matrix field equation:
\begin{equation}
  \label{genMuk}
  u^{I \prime\prime} + \left( \left[k^2 - \frac{a''}{a}\right] \delta^I_J
  + {\M^I}_J a^2 \right) u^J = 0 ,
\end{equation}
where a prime denotes derivatives with respect to $\eta$.
Since the mass term ${\M^I}_J$ is not in general diagonal, 
Eq. \eref{genMuk} does not
necessarily factorize into $\N$ independent equations for the $u^I$.  
Instead, the interactions described by ${\M^I}_J$ typically 
couple the scalar fields to one another.

In the single-field scenario, 
the mass-matrix ${\M^I}_J$ can be expressed as a combination
of slow-roll parameters. It can therefore be 
neglected to leading order in slow-roll.
To arrive at a similar result in the multi-field case, 
one can express $\M_{IJ}$ directly in terms of the slow-roll matrices 
\eref{epsilonIJ}--\eref{threept:etatilde}. It follows that 
\begin{equation}
\label{MIJslowroll}
  \frac{1}{H^2}\M_{IJ} = 3 \tilde{\eta}_{IJ} - 6 \epsilon_{IJ} -
  \epsilon \epsilon_{IJ} + \epsilon \eta_{IJ} 
\end{equation}
and Eq. \eref{MIJslowroll} 
generalizes the familiar result from single-field inflation.
Since we are only interested in leading order effects, we can 
consistently neglect the slow-roll contributions 
arising from $\M_{IJ}$ and consider only terms of $\Or(1)$. 
This is the standard approximation which is invoked when
estimating the amplitude and spectral index of inflationary fluctuations to
lowest order in slow-roll \cite{stewart-lyth,liddle-lyth},
and when calculating the leading-order effects of non-gaussianity
\cite{maldacena-nongaussian}. It results in the de Sitter--Mukhanov equation:
\begin{equation}
  u^{I \prime\prime} + \left( k^2 - \frac{a''}{a} \right) u^I =
  u^{I \prime\prime} + \left( k^2 - \frac{2}{\eta^2} \right) u^I = 0 .
\end{equation}

Applying the above simplification implies that we can work with the propagator
\begin{equation}
  \langle Q^I(x_1) Q^J(x_2) \rangle = \G^{IJ}
  G_{\star}(x_1,x_2) ,
\end{equation}
where $G_{\star}$ represents the conventional
scalar de Sitter Green's function for a massless scalar field
\cite{birrell-davies}:
\begin{equation}
  G_{\star}
  (x_1,x_2) =
  \frac{H^2}{2k^3} \times \left\{
  \begin{array}{ll}
    (1 - \imag k \eta_2)(1+ \imag k \eta_1) \e{-\imag k(\eta_1 - \eta_2)}
    & \quad \eta_1 > \eta_2 \\
    (1 - \imag k \eta_1)(1+ \imag k \eta_2) \e{\imag k(\eta_1 - \eta_2)}
    & \quad \eta_2 > \eta_1
  \end{array} \right.
\end{equation}
By taking the coincidence limit of the propagator, we obtain the familiar
power spectrum on large scales (where $k \rightarrow 0$): 
\begin{equation}
  \langle Q^I(\vect{k}_1) Q^J(\vect{k}_2) \rangle =
  (2\pi)^3 \delta(\vect{k}_1+ \vect{k}_2) \G^{IJ} \frac{H^2}{2k^3} ,
\end{equation}
and it therefore follows that 
\begin{equation}
\Delta^2_{\star} = \frac{H^2}{4\pi^2} .
\end{equation}

\subsection{The third-order theory}

The leading-order slow-roll term in the third-order action 
is given by\footnote{This expression should be compared, 
for example, with Eq. (3.6)
of \cite{maldacena-nongaussian}, to which it reduces in the single-field
case.}
\begin{eqnarray}
\label{S3}
  \nonumber
  \fl S_3 = \int \d t \, \dn{3}{x} \; a^3
  \Bigg( - \frac{1}{a^2} \G_{IJ} \dot{Q}^I \partial\psi
  \partial Q^J - \frac{1}{4H} \G_{MN} \dot{\phi}^M Q^N \G_{IJ} \dot{Q}^I
  \dot{Q}^J \\
  \mbox{} - \frac{1}{a^4} \frac{1}{4H} \G_{MN} \dot{\phi}^M Q^N \G_{IJ}
  \partial Q^I \partial Q^J \Bigg) ,
\end{eqnarray}
where $\psi$ was defined in \eref{threept:constr-def} and is given
to leading order by
\begin{equation}
  \label{threept:psi}
  \partial^2 \psi = - \frac{a^2}{2H} \G_{IJ} \dot{\phi}^I \dot{Q}^J .
\end{equation}
In exact analogy with the single-field calculation, it proves 
most convenient to
integrate by parts in the term containing $\partial\psi$.  
One then finds that 
\begin{eqnarray}
  \fl\nonumber
  - \int \d t \, \dn{3}{x} \; a \G_{IJ} \dot{Q}^I \partial \psi \partial
  Q^J = \\ \int \d t \, \dn{3}{x} \Bigg( -\frac{1}{2} aH \G_{IJ} \psi
  \partial Q^I \partial Q^J - \frac{1}{2} a \G_{IJ} \dot{\psi} \partial Q^I
  \partial Q^J + a \G_{IJ} \dot{Q}^I \psi \partial^2 Q^J \Bigg) .
\end{eqnarray}

Eq. \eref{threept:psi} may then be differentiated 
to determine $\dot{\psi}$. Keeping only leading-order terms, we find that 
\begin{equation}
\label{dotpsi}
  \dot{\psi} = - a^2 \G_{IJ} \dot{\phi}^I \partial^{-2} \dot{Q}^J -
  \frac{a^2}{2H} \G_{IJ} \dot{\phi}^I \partial^{-2} \ddot{Q}^J .
\end{equation}
It should be emphasized that \eref{dotpsi} should not be employed 
directly in the action \eref{S3} since it involves
second derivatives of the canonical field $Q^I$ and therefore changes
the order of the field equations. Instead, 
by neglecting subleading terms in slow-roll, one may 
use the first-order equation of motion 
\begin{equation}
  \frac{1}{a^3} \left.\frac{\delta L}{\delta Q^I}\right|_1 =
  - 3 H \dot{Q}_I - \ddot{Q}_I + \frac{1}{a^2} \partial^2 Q_I +
  \Or(\epsilon)
\end{equation}
to eliminate the second-derivative term
$\ddot{Q}^I$ in \eref{dotpsi}. It follows that 
\begin{equation}
\label{psidotA}  \fl
  \dot{\psi} = - a^2 \dot{\phi}^I \partial^{-2} \dot{Q}^I -
  \frac{a^2}{2H} \dot{\phi}^I \partial^{-2} \left(
  -\frac{1}{a^3} \left. \frac{\delta L}{\delta Q^I}\right|_1
  - 3H \dot{Q}_I + \frac{1}{a^2} \partial^2 Q_I \right) + \Or(\epsilon) .
\end{equation}
Substituting \eref{psidotA} into \eref{S3} and performing 
a further integration by parts where necessary then results in the equivalent
third-order action
\begin{eqnarray}
  \nonumber\fl
  S_3 = \int \d t \, \dn{3}{x} \; \Bigg( - \frac{a^3}{4H} \dot{\phi}^J
  Q_J \dot{Q}_I \dot{Q}^I - \frac{a^3}{2H} \dot{\phi}^J \partial^{-2} \dot{Q}_J
  \dot{Q}_I \partial^2 Q^I \\
  \label{threept:vertex}
  \mbox{} + \left.\frac{\delta L}{\delta{Q}_I}\right|_1 \left[
  \frac{1}{4H} \dot{\phi}^J \G_{IJ} \partial^{-2} ( Q^K \partial^2 Q_K ) -
  \frac{1}{8H} \dot{\phi}^J \G_{IJ} Q^K Q_K \right] \Bigg)  .
\end{eqnarray}

The last term in \eref{threept:vertex}, 
which is proportional to the first-order equation of motion
$\delta L/\delta Q^I|_1$, is familiar from the single-field calculation, and
represents a field-redefinition.  It can be eliminated from the action by
transforming the fields $Q^I$ to new fields $\Q^I$, which satisfy
\begin{equation}
  \label{threept:redef-explicit}
  \fl
  Q^I = \Q^I - F^I(Q) =
  \Q^I - \left( \frac{1}{4H} \dot{\phi}^I \partial^{-2}
  \left(  \Q^J \partial^2 \Q_J \right) - \frac{1}{8H} \dot{\phi}^I
  \Q^J \Q_J \right) ,
\end{equation}
where
it makes no difference to leading order whether we write $Q^J$ or $\Q^J$
in the quadratic terms.  Since we are keeping only terms up to and including
third order in $Q$, this redefinition has no effect on the
interactions described in \eref{threept:vertex}, and we may freely
set $Q^I = \Q^I$ there.  However, the redefinition does modify the
quadratic part of the action, Eq. \eref{S2}, which transforms as
\begin{equation}
  S_2[Q] \mapsto S_2[\Q] -
  \int \d t \, \dn{3}{x} \; \G_{IJ} \left. \frac{\delta L}{\delta Q^I}
  \right|_1 F^J .
\end{equation}
The extra term here exactly suffices to cancel the interactions in
\eref{threept:vertex} which were proportional to the first-order equations
of motion, $\delta L/\delta Q^I|_1$.

Although this field redefinition is extremely convenient for the purposes
of calculation,
we are ultimately interested in the correlators of $Q^I$ and
not $\Q^I$.  The correlators are related via the standard prescription
\begin{eqnarray}
  \fl\nonumber
  \langle Q^I Q^J Q^K \rangle = \int \fmeasure{Q^M} \, Q^I Q^J Q^J
  \e{\imag S[Q]} \\ \label{threept:wick}
  \lo{\simeq} \int \fmeasure{\Q^M} \, \left( \Q^I \Q^J \Q^K +
  F^I \Q^J \Q^K + \mbox{cyclic} \right)
  \e{\imag S[\Q]} + \Or(\epsilon) ,
\end{eqnarray}
which is just Wick's theorem.  This allows us to add in the contribution of
the redefined terms in \eref{threept:redef-explicit} at the end of the
calculation.

In the present calculation, the very complicated field redefinition found
by Maldacena (displayed in Eq. (3.8) of Ref. \cite{maldacena-nongaussian})
does not appear.  This is because such a redefinition of fields
is almost exactly
equivalent to a translation from the comoving to the uniform-curvature
gauge.  In this paper, we have worked in the uniform-curvature gauge from the
outset, which results in a considerable simplification even for the example
of a single field.  In some sense,
this effect demonstrates that the uniform-curvature gauge is especially
convenient for calculations of this sort: in the single-field case, one is
free to work equally in the comoving gauge (as in
Refs. \cite{maldacena-nongaussian,seery-lidsey}) or the
uniform-curvature gauge, but
in order to actually perform the functional integrals implicit in
the three-point correlator $\langle \zeta\zeta\zeta \rangle$, one must
eventually return to the uniform-curvature gauge, except 
for an anomalous field redefinition related to the effect of
the $\ddot{Q}^I$ term.

It is interesting to note that if we had
na\"{\i}vely kept $\ddot{Q}^I$ in the
action, instead of replacing it with the first-order equation of motion, 
we would have omitted the contribution of the quadratic terms in
\eref{threept:redef-explicit}.  This may seem paradoxical, but, as
we pointed out above, the appearance of $\ddot{Q}^I$ in the action changes the
order of the field equations, and consequently their dynamical behaviour.
We wish to retain only perturbations around the branch of solutions described
by \eref{intro:action}, and no others.  For this reason, we are obliged to
eliminate any second- or higher-derivative terms which appear in the
slow-roll expansion of the action for the $Q^I$, and we achieve this 
by employing the background
attractor behaviour of the second-order action to select the correct
solution around which we perturb.

\subsection{Calculating the three-point function}

We are now in a position to calculate the three-point function
$\langle \Q^I(\vect{k}_1) \Q^J(\vect{k}_2) \Q^K(\vect{k}_3) \rangle$
corresponding to the action \eref{threept:vertex}, where we work with
the redefined field $\Q^I$, defined in 
Eq. \eref{threept:redef-explicit},  
so that the
$\delta L/\delta Q_I|_1$ terms are not present.  A detailed description of how
the three-point calculation is carried out has been presented elsewhere
\cite{maldacena-nongaussian,seery-lidsey}, so here 
we merely record the results.

\begin{itemize}
  \item The first interaction to consider is given by 
   \begin{equation}
     \int \d \eta \, \dn{3}{x} \; \left( - \frac{a^2}{4H} \dot{\phi}^J
     \Q_J \Q^{\prime}_I \Q^{\prime I} \right) ,
  \end{equation}
  where we have converted the integral to conformal time. 
  This gives a contribution to $\langle \Q^I \Q^J \Q^K \rangle$,
  after translation to Fourier space, of the form 
  \begin{equation}
    \fl
    - \imag (2\pi)^3 \delta( \sum_i \vect{k}_i ) \frac{H^3}{4}
    \frac{1}{\prod_i 2 k_i^3} \dot{\phi}^I \G^{JK} \int \d \eta \;
    k_2^2 k_3^2 (1 - \imag k_1 \eta) \e{\imag k_t \eta} +
    \mbox{perms} ,
  \end{equation}
  where $k_t = k_1 + k_2 + k_3$ is the total (scalar) momentum. 
  The integral can be evaluated after Wick rotation onto the 
  positive imaginary axis and it then follows that this contribution 
  is equivalent to 
  \begin{equation}
    (2\pi)^3 \delta ( \sum_i \vect{k}_i ) \frac{H^3_{\ast}}{4}
    \frac{1}{\prod_i 2 k_i^3} \left[
    \sum_{\mbox{\scriptsize perms}} \dot{\phi}^I_{\ast}
    \G^{JK} \left( - \frac{k_2^2
    k_3^2}{k_t} - \frac{k_1 k_2^2 k_3^2}{k_t^2} \right) \right] ,
  \end{equation}
  where an asterisk `$\ast$' denotes evaluation at horizon crossing.
  The sum in this and subsequent expressions is over all ways of
  simultaneously rearranging the indices $I$, $J$ and $K$ and the
  momenta $k_1$, $k_2$ and $k_3$, such that the relative positioning of
  the $k$'s is respected.  (In other words, when exchanging indices $I$ and
  $J$, for example, one should also exchange $k_1$ and $k_2$, and so on.)
  
  \item The second relevant interaction is
  \begin{equation}
    \int \d \eta \, \dn{3}{x} \; \left( - \frac{a^2}{2H}
    \dot{\phi}^J \partial^{-2} \Q^{\prime}_J \Q^{\prime}_I \partial^2 \Q^I
    \right) ,
  \end{equation}
  where again we have rewritten the integral in terms of conformal time.
  After translating to Fourier space and Wick rotating the resulting
  integral, we have that 
  \begin{equation}
    (2\pi)^3 \delta( \sum_i \vect{k}_i ) \frac{H^3_{\ast}}{2}
    \frac{1}{\prod_i 2k_i^3} \left[
    \sum_{\mbox{\scriptsize perms}} \dot{\phi}^I_{\ast}
    \G^{JK} \left( - \frac{k_2^2
    k_3^2}{k_t} - \frac{k_2^2 k_3^3}{k_t^2} \right) \right] .
  \end{equation}
\end{itemize}

Once the field redefinition terms in \eref{threept:vertex} have been 
introduced back into the three-point function, using
\eref{threept:wick}, one finds that
\begin{equation}
  \label{threept:result}
  \langle Q^I(\vect{k}_1) Q^J(\vect{k}_2) Q^K(\vect{k}_3) \rangle =
  (2\pi)^3 \delta(\sum_i \vect{k}_i) \frac{4\pi^4}{\prod_i k_i^3}
  |\Delta_{\star}^2|^2 
  \A^{IJK}(k_1,k_2,k_3) ,
\end{equation}
where the spectrum $\Delta_{\star}^2$ is to be evaluated at horizon
crossing, and $\A^{IJK}$ is a momentum-dependent function given by 
\begin{eqnarray}
  \nonumber\fl
  \A^{IJK}(k_1,k_2,k_3) =
  \sum_{\mbox{\scriptsize perms}}
  \frac{\dot{\phi}^I_{\ast}}{4H_{\ast}}
  \G^{JK} \left( - 3 \frac{k_2^2 k_3^2}{k_t} -
  \frac{k_2^2 k_3^2}{k_t^2}(k_1 + 2 k_3) + \frac{1}{2} k_1^3 -
  k_1 k_2^2 \right) \\ \label{threept:a}
  \lo{=} \sum_{\mbox{\scriptsize perms}}
  \frac{1}{2\sqrt{2}} \epsilon^{I}_{\ast}
   \G^{JK}  \left( - 3 \frac{k_2^2 k_3^2}{k_t} -
  \frac{k_2^2 k_3^2}{k_t^2}(k_1 + 2 k_3) + \frac{1}{2} k_1^3 -
  k_1 k_2^2 \right)
\end{eqnarray}
This is of order a slow-roll parameter, so (as we expect)
the non-gaussianity will not be large when $|\epsilon^{IJ}| \ll 1$.
Eqs. (\ref{threept:result})--(\ref{threept:a}) 
represent the main result of this paper. 

The Feynman integrals that lead to $\langle Q^I Q^J Q^K \rangle$ 
are insensitive to the field modes' behaviour deep inside the horizon and 
long after the modes have passed outside the horizon. The
dominant contribution arises from fluctuations around the epoch of 
horizon crossing \cite{weinberg-corrl}.  For this reason,
\eref{threept:result} represents the primordial three-point correlation
among the scalars $Q^I$ a short time after horizon crossing.
In particular, in making the estimate
\eref{threept:result}--\eref{threept:a}, we have assumed that the three
$\vect{k}$-modes have roughly comparable wavenumbers, so that they cross
the horizon at similar epochs.

Eqs.~\eref{threept:result}--\eref{threept:a} do not by themselves
represent the primordial non-gaussianity; instead, they only describe cubic
interactions among the $Q^I$ at horizon crossing.
In order to use Eqs.~\eref{threept:result}--\eref{threept:a}
to calculate the primordial non-gaussianity in the
curvature perturbation, it is necessary to use the general formalism 
outlined in Section~\ref{sec:mfield} to assemble these initial 
non-gaussianities into the actual non-gaussianities that are in principle 
observable in the CMB.  

In the following sections, we consider two specific scenarios in which
we can carry out this detailed assembly explicitly.
We first verify that our results reduce to 
the well-known result of single-field inflation. We 
then proceed in Section \ref{sec:assisted} investigate the 
multiple-field analogue of the single-field power-law model
known as assisted inflation \cite{liddle-mazumdar}.

\section{Reduction to single-field case}
\label{sec:single}

In this section we verify that 
Eq.~\eref{threept:result} reduces to the result 
of Maldacena \cite{maldacena-nongaussian} for the case of inflation 
driven by a single scalar field.  We work with the large-scale
expression \eref{mfield:zeta}, which requires an explicit expression for the
number of e-folds, $N$, as a function of the field.
For a single field $\phi$ with perturbation $Q$ the integrated e-folding rate
is given by \eref{intro:ndef}:
\begin{equation}
  N = \int H \, \d t = - \frac{1}{\epsilon} \ln H
\end{equation}
whenever $\epsilon$ is approximately constant over the range of e-folds under
consideration.  Therefore, the derivatives of $N$, which are needed to 
determine $\zeta$, are
\begin{equation}
  \label{single:nderivs}
  N_{,\phi} = - \frac{1}{\epsilon} \frac{H_{,\phi}}{H} ,
  \qquad \qquad   N_{,\phi \phi} = - \frac{1}{\epsilon} 
  \frac{H_{,\phi \phi}}{H} +
  \frac{1}{\epsilon} \left( \frac{H_{,\phi}}{H} \right)^2 .
\end{equation}
Recalling that for single field models, the slow-roll
parameters can be expressed as 
\begin{equation}
  \epsilon = 2 \left( \frac{H_{,\phi}}{H} \right)^2  ,
  \qquad \qquad
  \eta = - \frac{\ddot{\phi}}{H\dot{\phi}} = 2 \frac{H_{,\phi \phi}}{H} ,
\end{equation}
we may write $\zeta$ as
\begin{equation}
  \zeta = - \frac{1}{\sqrt{2\epsilon}} Q + \left( \frac{1}{4} -
  \frac{\eta}{4\epsilon} \right) Q^2 .
\end{equation}

As expected, to lowest-order this implies 
that the spectra are related by the well-known result  
\cite{liddle-lyth-paper,liddle-lyth}
\begin{equation}
  \langle \zeta(\vect{k}_1) \zeta(\vect{k}_2) \rangle =
  \frac{1}{2\epsilon} \langle Q(\vect{k}_1) Q(\vect{k}_2) \rangle ,
\end{equation}
with $\epsilon$ evaluated at the moment of horizon crossing.
A similar expression holds for the three-point function.
It follows from \eref{threept:wick} that the $\langle \zeta\zeta\zeta
\rangle$ and $\langle QQQ \rangle$ correlators are related via
\begin{eqnarray}
\label{connect}
  \fl\nonumber
  \langle \zeta(\vect{k}_1) \zeta(\vect{k}_2) \zeta(\vect{k}_3) \rangle
  = - \frac{1}{(2\epsilon)^{3/2}} \langle Q(\vect{k}_1)
  Q(\vect{k}_2) Q(\vect{k}_3) \rangle \\ \label{single:redef}
  \mbox{} + \frac{1}{2\epsilon} \left(\frac{1}{4} - \frac{\eta}{4\epsilon}
  \right) \langle Q(\vect{k}_1) Q(\vect{k}_2) [Q \star
  Q](\vect{k}_3) \rangle + \mbox{perms} ,
\end{eqnarray}
where (as before) $\star$ denotes a convolution product.
Eq. (\ref{connect}) 
implies that the \emph{connected} part of the $\zeta$ three-point function is
related to the connected $Q$ three-point function by
\begin{equation}
  \langle \zeta(\vect{k}_1) \zeta(\vect{k}_2) \zeta(\vect{k}_3) \rangle_c =
  - \frac{H^3_{\ast}}{\dot{\phi}_{\ast}^3} \langle
  Q(\vect{k}_1) Q(\vect{k}_2) Q(\vect{k}_3) \rangle_c
\end{equation}
where a subscript $c$ denotes the connected part of a correlation
function.  However, the prescription of \eref{single:redef}
means that we must mix some disconnected pieces
with this connected component, which are described by the four-term
correlators $\langle QQQQ \rangle$.  These are no more than the
superhorizon parts of the gauge transformation between the comoving and
uniform curvature gauges,  which were derived by Maldacena
\cite{maldacena-nongaussian} and found to have the form
\begin{equation}
  (2\pi)^3 \delta(\sum_i \vect{k}_i) \frac{H^4}{4} \left(
  \frac{1}{2} \frac{H^4}{\dot{\phi}^4} \frac{\ddot{\phi}}{H\dot{\phi}} +
  \frac{1}{4} \frac{H^2}{\dot{\phi}^2} \right) \frac{1}{k_2^2 k_3^3} +
  \mbox{perms} .
\end{equation}
Thus, the complete three-point function which we have calculated 
via this method is
\begin{equation}
  \langle \zeta(\vect{k}_1) \zeta(\vect{k}_2) \zeta(\vect{k}_3) \rangle =
  (2\pi)^3 \delta(\sum_i \vect{k}_i) \frac{H_{\ast}^8}{\dot{\phi}_{\ast}^4}
  \frac{1} {\prod_i 2 k_i^3} \A ,
\end{equation}
where $\A$ is given by
\begin{equation}
  \A = \frac{\dot{\phi}_{\ast}^2}{H_{\ast}^2} \frac{4}{k_t}
  \sum_{i>j} k_i^2 k_j^2 + \left( 2 \frac{\ddot{\phi}_{\ast}}{H_{\ast}
  \dot{\phi}_{\ast}} +
  \frac{1}{2} \frac{\dot{\phi}_{\ast}^2}{H_{\ast}^2} \right) \sum_i k_i^3 +
  \frac{1}{2} \frac{\dot{\phi}_{\ast}^2}{H_{\ast}^2} \sum_{i \neq j} k_i k_j^2 ,
\end{equation}
in exact agreement with Maldacena's result.

\section{Assisted inflation}
\label{sec:assisted}

In this section we calculate the three-point 
correlation function for a multiple-field model of inflation
known as assisted inflation \cite{liddle-mazumdar}. 
In this scenario, a set of
$\N$ scalar fields, each with an exponential potential,
can conspire to drive a phase of accelerated expansion,  
even if the individual potential for each field is too steep
to support inflation. The potentials are given by 
\begin{equation}
\label{potI}
  V_I(\phi^I) = V_0 \exp \left( - \sqrt{\frac{2}{p^I}} \phi^I \right) ,
\end{equation}
where $V_0$ is a constant that determines the overall scale of the model, 
and $p^I$ represent coupling constants.\footnote{The summation convention for 
indices $I$, \ldots, $J$ is suspended in this section.}  In the single 
field model, $p^1>1$ is required for inflation. 
The combined effective potential is given by $V = \sum_I V_I$.

In the spatially flat FRW cosmology, 
the system of fields (\ref{potI}) admit a late-time attractor 
solution \cite{malik-wands} characterized by
\begin{equation}
  \phi^I = \sqrt{\frac{p^I}{p^1}} \phi^1 + \alpha^I ,
\end{equation}
for some constants $\alpha^I$ and any choice of reference field, which we
have arbitrarily identified as $\phi^1$. At the classical level, 
the existence of this attractor solution implies that the model
is dynamically equivalent to that of a single scalar field with an 
exponential potential characterized by a 
coupling constant $\tilde{p} = \sum_I p^I$. 

To proceed, we need to determine 
the number of e-folds in terms of the scalar 
field values. 
Following a standard calculation \cite{lyth-riotto}, we can express $\d N$ as
\begin{equation}
  \d N = - H \frac{\d \phi_I}{\dot{\phi}_I} ,
\end{equation}
for any field $\phi_I$.  
When slow-roll is valid, we can also make use of the approximate 
field equations $3H \dot{\phi}_I + \d V_I/\d\phi_I = 0$, such that 
\begin{equation}
  \label{assist:n}
  \d N \simeq \frac{3H^2}{\d V_1/\d \phi_1} \d \phi_1 \simeq 
  \frac{V}{\d V_1/\d \phi_1}
  \d \phi_1 = \sum_I \frac{V_I}{\d V_I / \d\phi_I} \, \d \phi_I .
\end{equation}
The first and second derivatives of $N$ now follow immediately 
from Eq. (\ref{assist:n}): 
\begin{eqnarray}
  \frac{\d N}{\d\phi_I} = \frac{V_I}{\d V_I/\d\phi_I}
  \nonumber \\
  \frac{\d^2 N}{\d\phi_I^2} = 1 - V_I \frac{\d^2 V_I}{\d\phi_I^2} 
  \left( \frac{\d V_I}{\d\phi_I} \right)^{-2} .
\end{eqnarray}
Moreover, for exponential potentials it follows that 
\begin{equation}
  \frac{V_I}{\d V_I/\d \phi_I} = - \sqrt{\frac{p^I}{2}} , \qquad
  \frac{\d^2 V_I/\d \phi_I^2}{\d V_I/\d \phi_I} = - \sqrt{\frac{2}{p^I}} .
\end{equation}
Hence, the second derivative of $N$ vanishes and 
the only remaining contribution to the 
total curvature perturbation \eref{mfield:zeta}
is provided by the linear derivative term,  which simplifies to
\begin{equation}
  \zeta = - \sum_I \sqrt{\frac{p^I}{2}} Q^I .
\end{equation}

At the level of the two-point correlation function, this 
implies that 
\begin{equation}
  \langle \zeta(\vect{k}_1) \zeta(\vect{k}_2) \rangle =
  \sum_{I, J} \frac{\sqrt{p^I p^J}}{2}
  \langle Q^I Q^J \rangle =
  (2\pi)^3 \delta(\vect{k}_1 + \vect{k}_2) \tilde{p} \frac{H^2_{\ast}}{4k_1^3} 
\end{equation}
(with an asterisk `$\ast$' indicating evaluation at horizon exit, as usual)
and this is precisely the spectrum that would 
arise for a single-field model with an exponential potential 
characterized by $\tilde{p}$, in agreement with the classical observation
that assisted inflation is dynamically equivalent to such a model
\cite{malik-wands}.  

On the other hand, we find that the three-point function is given by 
\begin{equation}
  \fl
  \langle \zeta(\vect{k}_1) \zeta(\vect{k}_2) \zeta(\vect{k}_2) \rangle =
  - \sum_{I, J, L} \sqrt{\frac{p^I p^J p^K}{8}} (2\pi)^3 \delta(\sum_i
  \vect{k}_i) \frac{H_{\ast}^4}{4} \frac{1}{\prod_i k_i^3} \A^{IJK}(
  k_1, k_2, k_3) .
\end{equation}
In order to correctly estimate this amplitude, we need an expression
for the slow-roll parameters. We take
\begin{equation}
  \epsilon^I = \frac{\sqrt{p^I}}{\tilde{p}} .
\end{equation}
Thus, $\tr \epsilon^{IJ} = 1/\tilde{p}$, which coincides with the slow-roll
parameter in the classically equivalent single-field model.
After some algebra it can be shown that this is equivalent to 
\begin{equation}
  \label{assist:result}
  \fl
  \langle \zeta(\vect{k}_1) \zeta(\vect{k}_2) \zeta(\vect{k}_2) \rangle =
  (2\pi)^3 \delta(\sum_i \vect{k}_i) \frac{H_{\ast}^4}{\prod_i 2k_i^3}
  \frac{1}{2} \tilde{p} \left( \frac{4}{k_t} \sum_{i>j} k_i^2 k_j^2 -
  \frac{1}{2} \sum_i k_i^3 + \frac{1}{2} \sum_{i \neq j} k_i k_j^2 \right) ,
\end{equation}
which is identical to the equivalent single-scalar model evaluated with
Maldacena's single-field formula. 

This reduction to the single-field result
illustrates a general feature of multiple-field models. In what amounts
to the familiar Gram--Schmidt procedure, one can always make a rotation in
field space so that one distinguished field (say $Q^1$) lies along the
adiabatic direction, and the remaining $\N - 1$ fields describe isocurvature
fluctuations around it \cite{gordon-wands,nibbelink-vantent}.
These isocurvature fluctuations do not contribute to the spectrum or
bispectrum. Therefore, quite generally, one expects the resultant three-point
function to be given by Maldacena's single-field result, constructed using
the field-space trajectory of the adiabatic field.

One should not understand this to mean that the formalism outlined in this
paper is unnecessary. On the contrary, although the rotation in field
space which produces an adiabatic field and $\N-1$ isocurvature fields
can be performed explicitly in simple cases \cite{gordon-wands}, for a
completely general multiple-field model, the most convenient way to account
for the effect of this rotation (at least on large scales)
is to use exactly the non-linear extension \cite{lyth-malik} of the
Sasaki--Stewart $\delta N$ formalism, which was described in
Section~\ref{sec:zeta}.
In this formalism it will always be necessary to `prime' the superhorizon
evolution using what amounts to an initial condition,
describing the non-gaussianity produced at horizon crossing. 
Eq. \eref{threept:a} constitutes exactly the necessary initial condition.

\section{Conclusions}
\label{sec:conclusions}

In this paper, we have calculated the three-point function 
evaluated just after horizon exit that arises 
from a set of $\N$ scalar field fluctuations $Q^I$ coupled to gravity
during inflation. The result is given by Eq. (\ref{fnl}), 
where the momentum-dependent contribution $\A^{IJK}$ is given in 
Eqs. (\ref{threept:result})--(\ref{threept:a}). 
This three-point function measures the intrinsic non-gaussianity
produced at horizon exit in the coupled scalar--gravity system,
and is important since it provides the initial condition needed 
for a calculation of the superhorizon evolution of the non-gaussianity 
in the inflationary density perturbation
\cite{lyth-rodriguez-a}. Such
superhorizon physics describes how the curvature perturbation $\zeta$,
which is used to calculate observable microwave background anisotropies
via the Sachs--Wolfe effect, evolves outside the Hubble radius.
Even though the initial non-gaussianity we have calculated, which is set
at horizon exit, is typically small, the total non-gaussianity of $\zeta$
after superhorizon evolution can be large. For example, this can occur
in curvaton-like scenarios \cite{boubekeur-lyth} or where there is
cross-correlation between the adiabatic and isocurvature modes
\cite{bartolo-a}.

As a result, 
in some circumstances our expression could be used to verify that the
non-gaussian effects generated at horizon crossing by microphysical
processes are subdominant to the superhorizon, isocurvature-driven
evolution of $\zeta$, as described in \cite{lyth-rodriguez-a}.  However,
more generally, both of these effects will be important and our
result contributes in an essential way to the primordial non-gaussianity
produced by inflation.

It is important to note that \eref{threept:result}--\eref{threept:a}
describe the non-linearities among the $Q^I$ at horizon crossing.
The details of this were recently explored by Weinberg \cite{weinberg-corrl},
who showed that in a large class of multi-field models such integrals
are dominated by the epoch of horizon-crossing, even beyond tree-level in
perturbation theory. Provided one evaluates the slow-roll parameters when
the $\vect{k}$-modes in question crossed the horizon,
Eq. \eref{threept:a} gives the relevant non-linearities. Of course, both
the $Q^I$ and $\zeta$ continue to evolve on superhorizon scales.
For this reason one must assemble the $\langle QQQ \rangle$ correlators
into $\langle \zeta\zeta\zeta \rangle$ correlators to follow the
time dependence correctly when outside the Hubble radius. The same
conclusion was reached by Rigopoulos, Shellard \& van Tent
\cite{rigopoulos-shellard-vantent-b}.

We do not address the question of what happens when one $\vect{k}$-mode
is squeezed, so that $k_1 \ll k_2, k_3$ (say). In this case the mode
corresponding to $\vect{k}_1$ is pushed outside the horizon much earlier
than the other two, since it is comparatively larger. In the single-field
case, where perturbations ``freeze in'' outside the horizon owing to the
constancy of $\zeta$, the three-point function factorizes into
the two-point function of $\vect{k}_2$ and $\vect{k}_3$ evaluated in
a background which takes into account the back-reaction of the frozen
$\vect{k}_1$ perturbation \cite{maldacena-nongaussian}.
In a multiple-field model, the $\vect{k}_1$ mode
will come to constitute part of the zero-momentum background when it is far
outside the horizon, but since $\zeta$ is no longer constant it is not
so straightforward to provide a quantitative description of its effect.
We leave this interesting question for future work.

The level of non-gaussianity arising from microphysics is typically 
proportional to a slow-roll parameter, 
$\epsilon^{1/2} \A^{IJK}$, 
and is therefore expected to be small
when fluctuations are close to scale-invariance.
Consequently, when isocurvature modes drive
a very strong superhorizon evolution of $\zeta$, and
thereby source considerable non-gaussianity on very large scales, 
it seems likely that the generic situation 
will match that described by Lyth \& Rodriguez \cite{lyth-rodriguez-a},
although the only model we are aware of where this can be verified in
detail is the curvaton scenario.
On the other hand, if the superhorizon
contribution never becomes large, the non-gaussian effects we have 
considered will be important,   
in which case they can not be neglected. This
is the situation that arises, for example, in the assisted inflationary
scenario considered in Section~\ref{sec:assisted}, which is the 
canonical model where the superhorizon piece is entirely absent.
It is worth emphasizing that the three-point function can be calculated 
once the evolution of the unperturbed background cosmology, as 
parametrized by $N(\phi^I )$, has been determined.  
Our result therefore complements 
that of \cite{lyth-rodriguez-a}, and implies that whenever one has
enough information about the background 
dynamics to calculate the momentum-independent piece of $\fnl$,
there will be always be enough information to calculate the
momentum-dependent contribution as well.  
In this sense, no further work is required in order to 
calculate the full, momentum-dependent $\fnl$.

In general, the structure of the three-point function 
(\ref{threept:result})--(\ref{threept:a}) is sensitive 
to the target space metric $\G^{IJ}$.
A similar dependence arises in the two-point function, which 
is proportional to $\G^{IJ}$.  Indeed, 
neglecting numerical coefficients, the form of 
\eref{threept:a} is in some sense inevitable. Specifically,  
it must arise from a cubic combination of the $Q^A$'s which is
a singlet with respect to the target space metric.  
Since the $Q^A$ can only appear
contracted with the metric $\G_{IJ}$, it follows that to leading-order in
slow-roll, the only possible
combination must have the form $\G_{AB} \dot{\phi}^A Q^B \G_{IJ} Q^I Q^J$.
Such a combination leads to the
index structure appearing in \eref{threept:a}, there being no other
target space vector at lowest-order in slow-roll which can be
contracted with $G_{AB} Q^B$ to yield a singlet.

It is natural to speculate that an analogous result will hold for the general
$n$-th--order correlation function.  When $n$ is even ($n = 2m$) we expect
the leading order slow-roll contribution to arise from a scalar 
quantity containing $m$ contractions with $\G_{IJ}$:
\begin{equation}
  \label{threept:even}
  \langle \underbrace{Q^A Q^B \cdots Q^E Q^F}_{\mbox{\scriptsize{$2m$ copies}}}
  \rangle \sim
  \left( \frac{\dot{\phi}}{H} \right)^{2(m-1)}
  \underbrace{\G^{(AB} \cdots \G^{EF)}}_{\mbox{\scriptsize{$m$ copies}}} \, ,
\end{equation}
where we have neglected momentum-dependent factors
which may accompany each factor of the target space metric $\G^{AB}$.
When $n$ is odd, so that $n = 2m+1$, we anticipate instead that 
\begin{equation}
  \label{threept:odd}
  \langle \underbrace{Q^A Q^B Q^C \cdots Q^E Q^F}_
  {\mbox{\scriptsize{$2m+1$ copies}}} \rangle \sim
  \left( \frac{\dot{\phi}}{H} \right)^{2(m-1)}
  H^{-1} \dot{\phi}^{(A} \underbrace{\G^{BC} \cdots \G^{EF)}}_
  {\mbox{\scriptsize{$m$ copies}}} .
\end{equation}

We have restricted our attention to the case where the target-space 
metric is flat and independent of the scalar field values, $\G^{IJ} = 
\delta^{IJ}$. Although this Ansatz covers a very wide class of multi-field 
inflationary models, it would be interesting to go beyond this approximation 
(see, for example,
\cite{mukhanov-steinhardt,sasaki-stewart,nibbelink-vantent,van-tent}). 
Such an extension has been considered recently 
using a stochastic approach \cite{rigopoulos-shellard,
rigopoulos-shellard-vantent}, although in principle, more general 
scenarios of this type  
could be investigated by employing the formalism developed 
in the present  paper. For example, a specific point of interest concerns the
curvature of the scalar field manifold. In general, a non-trivial 
metric $\G_{IJ}(\phi)$ will result in a curvature tensor of the form 
\begin{equation}
  {\Omega^A}_{BCD} = {\omega^A}_{BD,C} - {\omega^A}_{BC,D} +
  {\omega^F}_{BD} {\omega^A}_{FC} - {\omega^F}_{BC} {\omega^A}_{FD} ,
\end{equation}
where ${\omega^A}_{BC}$ is the Levi-Civita connexion 
compatible with $\G_{IJ}$. This tensor is identically 
zero for the models 
considered in this paper, but is expected to arise in the 
expressions for the spectral tilt and other cosmological observables 
\cite{sasaki-stewart} in more general classes of models.
It is possible that the presence of curvature terms of this form may be 
sufficiently important to invalidate the flat target-space
analysis we have adopted. 
In this case, a larger primordial non-gaussianity could be generated.
However, since the curvature tensor ${\Omega^A}_{BCD}$ can not be too large 
if slow-roll is to be respected \cite{sasaki-stewart},
we anticipate that this will not be the case in general.

\ackn
DS is supported by PPARC. We would like to thank David Lyth and David Wands
for useful discussions.

\section*{\refname}
\providecommand{\href}[2]{#2}\begingroup\raggedright\endgroup


\begin{thebibliography}{10}

\bibitem{starobinsky}
A.~Starobinsky, {\it A new type of isotropic cosmological model without
  singularity},  {\em Phys. Lett. B} {\bf 91} (1980) 99--102.

\bibitem{guth}
A.~Guth, {\it Inflationary universe: {A} possible solution to the horizon and
  flatness problems},  {\em Phys. Rev. D} {\bf 23} (1981) 347--356.

\bibitem{albrecht-steinhardt}
A.~Albrecht and P.~Steinhardt, {\it Cosmology for {G}rand {U}nified {T}heories
  with radiatively induced symmetry breaking},  {\em Phys. Rev. Lett.} {\bf 48}
  (1982) 1220--1223.

\bibitem{hawking-moss}
S.~Hawking and I.~Moss, {\it Supercooled phase transitions in the very early
  universe},  {\em Phys. Lett. B} {\bf 110} (1982) 35.

\bibitem{linde}
A.~Linde, {\it A new inflationary universe scenario: {A} possible solution of
  the horizon, flatness, homogeneity, isotropy and primordial monopole
  problems},  {\em Phys. Lett. B} {\bf 108} (1982) 389--393.

\bibitem{lyth-riotto}
D.~Lyth and A.~Riotto, {\it Particle {P}hysics {M}odels of {I}nflation and the
  {C}osmological {D}ensity {P}erturbation},  {\em Phys. Rept.} {\bf 314} (1999)
  1--146, [\href{http://xxx.lanl.gov/abs/hep-ph/9807278}{{\tt
  hep-ph/9807278}}].

\bibitem{liddle-lyth}
A.~Liddle and D.~Lyth, {\em Cosmological Inflation and Large-Scale Structure}.
\newblock Cambridge University Press, Cambridge, 2000.

\bibitem{bennett-banday}
C.~Bennett, A.~Banday, K.~Gorski, G.~Hinshaw, P.~Jackson, P.~Keegstra,
  A.~Kogut, G.~Smoot, D.~Wilkinson, and E.~Wright, {\it Four year {COBE} {DMR}
  {C}osmic {M}icrowave {B}ackground observations: {M}aps and basic results},
  {\em Astrophys. J.} {\bf 464} (1996) L1--L4,
  [\href{http://xxx.lanl.gov/abs/astro-ph/9601067}{{\tt astro-ph/9601067}}].

\bibitem{wmap}
D.~Spergel, L.~Verde, H.~Peiris, E.~Komatsu, M.~Nolta, C.~Bennett, M.~Halpern,
  G.~Hinshaw, N.~Jarosik, A.~Kogut, M.~Limon, S.~Meyer, L.~Page, G.~Tucker,
  J.~Weiland, E.~Wollack, and E.~Wright, {\it First year {W}ilkinson
  {M}icrowave {A}nisotropy {P}robe ({W}{M}{A}{P}) observations: Determination
  of cosmological parameters},  {\em Astrophys. J. Suppl.} {\bf 148} (2003)
  175, [\href{http://xxx.lanl.gov/abs/astro-ph/0302209}{{\tt
  astro-ph/0302209}}].

\bibitem{bartolo-matarrese-review}
N.~Bartolo, E.~Komatsu, S.~Matarrese, and A.~Riotto, {\it Non-{G}aussianity
  from {I}nflation: {T}heory and {O}bservations},  {\em Phys. Rept.} {\bf 402}
  (2004) 103--266, [\href{http://xxx.lanl.gov/abs/astro-ph/0406398}{{\tt
  astro-ph/0406398}}].

\bibitem{maldacena-nongaussian}
J.~Maldacena, {\it Non-{G}aussian features of primordial fluctuations in single
  field inflationary models},  {\em JHEP} {\bf 0305} (2003) 013,
  [\href{http://xxx.lanl.gov/abs/astro-ph/0210603}{{\tt astro-ph/0210603}}].

\bibitem{seery-lidsey}
D.~Seery and J.~Lidsey, {\it Primordial non-gaussianity in single-field
  inflation},  {\em JCAP} {\bf 06} (2005) 003,
  [\href{http://xxx.lanl.gov/abs/astro-ph/0503692}{{\tt astro-ph/0503692}}].

\bibitem{creminelli}
P.~Creminelli, {\it On non-gaussianities in single-field inflation},  {\em
  JCAP} {\bf 0310} (2003) 003,
  [\href{http://xxx.lanl.gov/abs/astro-ph/0306122}{{\tt astro-ph/0306122}}].

\bibitem{lyth-rodriguez}
D.~Lyth and Y.~Rodriguez, {\it Non-gaussianity from the second-order
  cosmological perturbation},
  \href{http://xxx.lanl.gov/abs/astro-ph/0502578}{{\tt astro-ph/0502578}}.

\bibitem{lyth-rodriguez-a}
D.~Lyth and Y.~Rodriguez, {\it The inflationary prediction for primordial
  non-gaussianity},  \href{http://xxx.lanl.gov/abs/astro-ph/0504045}{{\tt
  astro-ph/0504045}}.

\bibitem{acquaviva-bartolo}
V.~Acquaviva, N.~Bartolo, S.~Matarrese, and A.~Riotto, {\it Second-{O}rder
  {C}osmological {P}erturbations from {I}nflation},  {\em Nucl. Phys. B} {\bf
  667} (2003) 119--148, [\href{http://xxx.lanl.gov/abs/astro-ph/0209156}{{\tt
  astro-ph/0209156}}].

\bibitem{calcagni-nongaussian}
G.~Calcagni, {\it Non-{G}aussianity in braneworld and tachyon inflation},
  \href{http://xxx.lanl.gov/abs/astro-ph/0411773}{{\tt astro-ph/0411773}}.

\bibitem{rigopoulos-shellard}
G.~Rigopoulos and E.~Shellard, {\it Non-linear inflationary perturbations},
  \href{http://xxx.lanl.gov/abs/astro-ph/0405185}{{\tt astro-ph/0405185}}.

\bibitem{rigopoulos-shellard-vantent}
G.~Rigopoulos, E.~Shellard, and B.~van Tent, {\it Non-linear perturbations in
  multiple-field inflation},
  \href{http://xxx.lanl.gov/abs/astro-ph/0504508}{{\tt astro-ph/0504508}}.

\bibitem{mcewen-hobson}
J.~McEwen, M.~Hobson, A.~Lasenby, and D.~Mortlock, {\it A high-significance
  detection of non-{G}aussianity in the {WMAP} 1-year data using directional
  spherical wavelets},  {\em Mon. Not. Roy. Astron. Soc.} {\bf 359} (2005)
  1583--1596, [\href{http://xxx.lanl.gov/abs/astro-ph/0406604}{{\tt
  astro-ph/0406604}}].

\bibitem{mukherjee-wang}
P.~Mukherjee and Y.~Wang, {\it Wavelets and {WMAP} non-{G}aussianity},
  \href{http://xxx.lanl.gov/abs/astro-ph/0402602}{{\tt astro-ph/0402602}}.

\bibitem{vielva-martinez-gonzalez}
P.~Vielva, E.~Martinez-Gonzalez, R.~Barreiro, J.~Sanz, and L.~Cayon, {\it
  Detection of non-{G}aussianity in the {WMAP} 1-year data using spherical
  wavelets},  {\em Astrophys. J.} {\bf 609} (2004) 22--34,
  [\href{http://xxx.lanl.gov/abs/astro-ph/0310273}{{\tt astro-ph/0310273}}].

\bibitem{larsen-wandelt}
D.~Larsen and B.~Wandelt, {\it A statistically robust 3-sigma detection of
  non-{G}aussianity in the {WMAP} data using hot and cold spots},
  \href{http://xxx.lanl.gov/abs/astro-ph/0505046}{{\tt astro-ph/0505046}}.

\bibitem{aghanim-kunz}
N.~Aghanim, M.~Kunz, P.~Castro, and O.~Forni, {\it Non-{G}aussianity:
  {C}omparing wavelet and {F}ourier based methods},  {\em Astron. Astrophys.}
  {\bf 406} (2003) 797--816,
  [\href{http://xxx.lanl.gov/abs/astro-ph/0301220}{{\tt astro-ph/0301220}}].

\bibitem{falk-rangarajan}
T.~Falk, R.~Rangarajan, and M.~Srednicki, {\it The {A}ngular {D}ependence of
  the {T}hree-{P}oint {C}orrelation {F}unction of the {C}osmic {M}icrowave
  {B}ackground {R}adiation as {P}redicted by {I}nflationary {C}osmologies},
  {\em Astrophys. J.} {\bf 403} (1993) L1,
  [\href{http://xxx.lanl.gov/abs/astro-ph/9208001}{{\tt astro-ph/9208001}}].

\bibitem{gangui-lucchin}
A.~Gangui, F.~Lucchin, S.~Matarrese, and S.~Mollerach, {\it The {T}hree-{P}oint
  {C}orrelation {F}unction of the {C}osmic {M}icrowave {B}ackground in
  {I}nflationary {M}odels},  {\em Astrophys. J.} {\bf 430} (1994) 447--457,
  [\href{http://xxx.lanl.gov/abs/astro-ph/9312033}{{\tt astro-ph/9312033}}].

\bibitem{alishahiha-silverstein}
M.~Alishahiha, E.~Silverstein, and D.~Tong, {\it D{BI} in the {S}ky},  {\em
  Phys. Rev. D} {\bf 70} (2004) 123505,
  [\href{http://xxx.lanl.gov/abs/hep-th/0404084}{{\tt hep-th/0404084}}].

\bibitem{arkani-hamed-creminelli}
N.~Arkani-Hamed, P.~Creminelli, S.~Mukhoyama, and M.~Zaldarriaga, {\it Ghost
  {I}nflation},  {\em JCAP} {\bf 0404} (2004) 001,
  [\href{http://xxx.lanl.gov/abs/hep-th/0312100}{{\tt hep-th/0312100}}].

\bibitem{bullock-primack}
J.~Bullock and J.~Primack, {\it Non-{G}aussian {F}luctuations and {P}rimordial
  {B}lack {H}oles from {I}nflation},  {\em Phys. Rev. D} {\bf 55} (1997)
  7423--7439, [\href{http://xxx.lanl.gov/abs/astro-ph/9611106}{{\tt
  astro-ph/9611106}}].

\bibitem{ivanov}
P.~Ivanov, {\it Non-linear metric perturbations and production of primordial
  black holes},  {\em Phys. Rev. D} {\bf 57} (1998) 7145--7154,
  [\href{http://xxx.lanl.gov/abs/astro-ph/9708224}{{\tt astro-ph/9708224}}].

\bibitem{bartolo-matarrese-inflation}
N.~Bartolo, S.~Matarrese, and A.~Riotto, {\it Enhancement of non-gaussianity
  after inflation},  {\em JHEP} {\bf 04} (2004) 006,
  [\href{http://xxx.lanl.gov/abs/astro-ph/0308088}{{\tt astro-ph/0308088}}].

\bibitem{bartolo-matarrese-fnl}
N.~Bartolo, S.~Matarrese, and A.~Riotto, {\it Gauge-{I}nvariant {T}emperature
  {A}nisotropies and {P}rimordial {N}on-{G}aussianity},  {\em Phys. Rev. Lett.}
  {\bf 93} (2004) 231301, [\href{http://xxx.lanl.gov/abs/astro-ph/0407505}{{\tt
  astro-ph/0407505}}].

\bibitem{bento-bertolami-a}
M.~Bento, O.~Bertolami, and P.~Sa, {\it {I}nflation from strings},  {\em Phys.
  Lett. B} {\bf 262} (1991) 11--17.

\bibitem{bento-bertolami-b}
M.~Bento, O.~Bertolami, and P.~Sa, {\it {I}nflation from strings. 2.
  {R}eheating and {B}aryogenesis},  {\em Mod. Phys. Lett. A} {\bf 7} (1992)
  911--920.

\bibitem{sakharov-khlopov}
A.~Sakharov and M.~Khlopov, {\it The nonhomogeneity problem for the primordial
  axion field},  {\em Phys. Atom. Nucl.} {\bf 57} (1994) 485--487.

\bibitem{sakharov-sokoloff}
A.~Saharov, D.~Sokoloff, and M.~Khlopov, {\it Large scale modulation of the
  distribution of coherent oscillations of primordial axion field in the
  {U}niverse},  {\em Phys. Atom. Nucl.} {\bf 59} (1996) 1005--1010.

\bibitem{enqvist-jokinen-a}
K.~Enqvist, A.~Jokinen, A.~Mazumdar, T.~Multamaki, and A.~V\"{a}ihk\"{o}nen,
  {\it Non-gaussianity from instant and tachyonic preheating},  {\em JCAP} {\bf
  0503} (2005) 010, [\href{http://xxx.lanl.gov/abs/hep-ph/0501076}{{\tt
  hep-ph/0501076}}].

\bibitem{enqvist-jokinen-b}
K.~Enqvist, A.~Jokinen, A.~Mazumdar, T.~Multamaki, and A.~V\"{a}ihk\"{o}nen,
  {\it Cosmological constraints on string scale and coupling arising from
  tachyonic instability},  \href{http://xxx.lanl.gov/abs/hep-th/0502185}{{\tt
  hep-th/0502185}}.

\bibitem{enqvist-vaihkonen}
K.~Enqvist and A.~V\"{a}ihk\"{o}nen, {\it Non-{G}aussian perturbations in
  hybrid inflation},  {\em JCAP} {\bf 0409} (2004) 006,
  [\href{http://xxx.lanl.gov/abs/hep-ph/0405103}{{\tt hep-ph/0405103}}].

\bibitem{enqvist-jokinen}
K.~Enqvist, A.~Jokinen, A.~Mazumdar, T.~Multamaki, and A.~V\"{a}ihk\"{o}nen,
  {\it Non-{G}aussianity from {P}reheating},  {\em Phys. Rev. Lett.} {\bf 94}
  (2005) 161301, [\href{http://xxx.lanl.gov/abs/astro-ph/0411394}{{\tt
  astro-ph/0411394}}].

\bibitem{bernardeau-uzan}
F.~Bernardeau and J.-P. Uzan, {\it Non-{G}aussianity in multi-field inflation},
   {\em Phys. Rev. D} {\bf 66} (2002) 103506,
  [\href{http://xxx.lanl.gov/abs/hep-ph/0207295}{{\tt hep-ph/0207295}}].

\bibitem{lyth-wands}
D.~Lyth and D.~Wands, {\it Generating the curvature perturbation without an
  inflaton},  {\em Phys. Lett. B} {\bf 524} (2002) 5--14,
  [\href{http://xxx.lanl.gov/abs/hep-ph/0110002}{{\tt hep-ph/0110002}}].

\bibitem{moroi-takahashi}
T.~Moroi and T.~Takahashi, {\it Effects of cosmological moduli fields on cosmic
  microwave background},  {\em Phys. Lett. B} {\bf 522} (2001) 215--221,
  [\href{http://xxx.lanl.gov/abs/hep-ph/0110096}{{\tt hep-ph/0110096}}].
  Erratum-ibid. B \textbf{539} 303 (2002).

\bibitem{enqvist-sloth}
K.~Enqvist and M.~Sloth, {\it Adiabatic {CMB} perturbations in {P}re-{B}ig
  {B}ang {S}tring {C}osmology},  {\em Nucl. Phys. B} {\bf 626} (2002) 395--409,
  [\href{http://xxx.lanl.gov/abs/hep-ph/0109214}{{\tt hep-ph/0109214}}].

\bibitem{sasaki-stewart}
M.~Sasaki and E.~Stewart, {\it A {G}eneral {A}nalytic {F}ormula for the
  {S}pectral {I}ndex of the {D}ensity {P}erturbations produced during
  {I}nflation},  {\em Prog. Theor. Phys.} {\bf 95} (1996) 71--78,
  [\href{http://xxx.lanl.gov/abs/astro-ph/9507001}{{\tt astro-ph/9507001}}].

\bibitem{liddle-mazumdar}
A.~Liddle and A.~Mazumdar, {\it Assisted inflation},  {\em Phys. Rev. D} {\bf
  58} (1998) 061301, [\href{http://xxx.lanl.gov/abs/astro-ph/9804177}{{\tt
  astro-ph/9804177}}].

\bibitem{barreiro-copeland}
T.~Barreiro, E.~Copeland, and N.~Nunes, {\it Quintessence arising from
  exponential potentials},  {\em Phys. Rev. D} {\bf 61} (2000) 127301,
  [\href{http://xxx.lanl.gov/abs/astro-ph/9910214}{{\tt astro-ph/9910214}}].

\bibitem{green-lidsey}
A.~Green and J.~Lidsey, {\it Generalized compactification and assisted dynamics
  of multiscalar field cosmologies},  {\em Phys. Rev. D} {\bf 61} (2000)
  067301, [\href{http://xxx.lanl.gov/abs/astro-ph/9907223}{{\tt
  astro-ph/9907223}}].

\bibitem{hwang-noh-multi}
J.~Hwang and H.~Noh, {\it Cosmological {P}erturbations with {M}ultiple
  {F}ields},  {\em Phys. Lett. B} {\bf 495} (2000) 277--283,
  [\href{http://xxx.lanl.gov/abs/astro-ph/0009268}{{\tt astro-ph/0009268}}].

\bibitem{nibbelink-vantent}
S.~G. Nibbelink and B.~van Tent, {\it Scalar perturbations during
  multiple-field slow-roll inflation},  {\em Class. Quant. Grav.} {\bf 19}
  (2002) 613--640, [\href{http://xxx.lanl.gov/abs/hep-ph/0107272}{{\tt
  hep-ph/0107272}}].

\bibitem{van-tent}
B.~van Tent, {\it Multiple-field inflation and the {CMB}},  {\em Class. Quant.
  Grav.} {\bf 21} (2004) 349--370,
  [\href{http://xxx.lanl.gov/abs/astro-ph/0307048}{{\tt astro-ph/0307048}}].

\bibitem{mfb}
V.~Mukhanov, H.~Feldman, and R.~Brandenberger, {\it Theory of {C}osmological
  {P}erturbations},  {\em Phys. Rept.} {\bf 215} (1992) 203.

\bibitem{wands-malik}
D.~Wands, K.~Malik, D.~Lyth, and A.~Liddle, {\it A new approach to the
  evolution of cosmological perturbations on large scales},  {\em Phys. Rev. D}
  {\bf 62} (2000) 043527, [\href{http://xxx.lanl.gov/abs/astro-ph/0003278}{{\tt
  astro-ph/0003278}}].

\bibitem{lyth-malik}
D.~Lyth, K.~Malik, and M.~Sasaki, {\it A general proof of the conservation of
  the curvature perturbation},  {\em JCAP} {\bf 0505} (2005) 004,
  [\href{http://xxx.lanl.gov/abs/astro-ph/0411220}{{\tt astro-ph/0411220}}].

\bibitem{boubekeur-lyth}
L.~Boubekeur and D.~Lyth, {\it Detecting a small perturbation through its
  non-{G}aussianity},  \href{http://xxx.lanl.gov/abs/astro-ph/0504046}{{\tt
  astro-ph/0504046}}.

\bibitem{gordon-wands}
C.~Gordon, D.~Wands, B.~Bassett, and R.~Maartens, {\it Adiabatic and entropy
  perturbations from inflation},  {\em Phys. Rev. D} {\bf 63} (2001) 023506,
  [\href{http://xxx.lanl.gov/abs/astro-ph/0009131}{{\tt astro-ph/0009131}}].

\bibitem{komatsu-spergel}
E.~Komatsu and D.~Spergel, {\it Acoustic {S}ignatures in the {P}rimary
  {M}icrowave {B}ackground {B}ispectrum},  {\em Phys. Rev. D} {\bf 63} (2001)
  063002, [\href{http://xxx.lanl.gov/abs/astro-ph/0005036}{{\tt
  astro-ph/0005036}}].

\bibitem{verde-wang}
L.~Verde, L.~Wang, A.~Heavens, and M.~Kamionkowski, {\it Large-scale structure,
  the cosmic microwave background and primordial non-gaussianity},  {\em Mon.
  Not. Roy. Astron. Soc.} {\bf 313} (2000) L141--L147,
  [\href{http://xxx.lanl.gov/abs/astro-ph/9906301}{{\tt astro-ph/9906301}}].

\bibitem{lyth-axion}
D.~Lyth, {\it Axions and inflation: {V}acuum fluctuations},  {\em Phys. Rev. D}
  {\bf 45} (1992) 3394--3404.

\bibitem{salopek-bond}
D.~Salopek and J.~Bond, {\it Nonlinear evolution of long-wavelength metric
  fluctuations in inflationary models},  {\em Phys. Rev. D} {\bf 42} (1990)
  3936--3962.

\bibitem{stewart-lyth}
E.~Stewart and D.~Lyth, {\it A more accurate analytic calculation of the
  spectrum of cosmological perturbations produced during inflation},  {\em
  Phys. Lett. B} {\bf 302} (1993) 171,
  [\href{http://xxx.lanl.gov/abs/gr-qc/9302019}{{\tt gr-qc/9302019}}].

\bibitem{birrell-davies}
N.~Birrell and P.~Davies, {\em Quantum fields in curved space}.
\newblock Cambridge University Press, Cambridge, 1982.

\bibitem{weinberg-corrl}
S.~Weinberg, {\it {Q}uantum {C}ontributions to {C}osmological {C}orrelations},
  \href{http://xxx.lanl.gov/abs/hep-th/0506236}{{\tt hep-th/0506236}}.

\bibitem{liddle-lyth-paper}
A.~Liddle and D.~Lyth, {\it The {C}old {D}ark {M}atter {D}ensity
  {P}erturbation},  {\em Phys. Rept.} {\bf 231} (1993) 1--105,
  [\href{http://xxx.lanl.gov/abs/astro-ph/9303019}{{\tt astro-ph/9303019}}].

\bibitem{malik-wands}
K.~Malik and D.~Wands, {\it Dynamics of {A}ssisted {I}nflation},  {\em Phys.
  Rev. D} {\bf 59} (1999) 123501,
  [\href{http://xxx.lanl.gov/abs/astro-ph/9812204}{{\tt astro-ph/9812204}}].

\bibitem{bartolo-a}
N.~Bartolo, S.~Matarrese, and A.~Riotto, {\it {N}on-{G}aussianity from
  inflation},  {\em Phys. Rev. D} {\bf 65} (2002) 103505,
  [\href{http://xxx.lanl.gov/abs/hep-ph/0112261}{{\tt hep-ph/0112261}}].

\bibitem{rigopoulos-shellard-vantent-b}
G.~Rigopoulos, E.~Shellard, and B.~van Tent, {\it Large {N}on-{G}aussianity in
  {M}ultiple-{F}ield {I}nflation},
  \href{http://xxx.lanl.gov/abs/astro-ph/0506704}{{\tt astro-ph/0506704}}.

\bibitem{mukhanov-steinhardt}
V.~Mukhanov and P.~Steinhardt, {\it Density {P}erturbations in {M}ultifield
  {I}nflationary {M}odels},  {\em Phys. Lett. B} {\bf 422} (1998) 52--60,
  [\href{http://xxx.lanl.gov/abs/astro-ph/9710038}{{\tt astro-ph/9710038}}].

\end{thebibliography}
\end{document}